# A Closely-Packed System of Low-Mass, Low-Density Planets Transiting Kepler-11

Jack J. Lissauer[1], Daniel C. Fabrycky[2], Eric B. Ford[3], William J. Borucki[1], Francois Fressin[4], Geoffrey W. Marcy[5], Jerome A. Orosz[6], Jason F. Rowe[7], Guillermo Torres[4], William F. Welsh[6], Natalie M. Batalha[8], Stephen T. Bryson[1], Lars A. Buchhave[9], Douglas A. Caldwell[7], Joshua A. Carter[4], David Charbonneau[4], Jessie L. Christiansen[7], William D. Cochran[10], Jean-Michel Desert[4], Edward W. Dunham[11], Michael N. Fanelli[12], Jonathan J. Fortney[2], Thomas N. Gautier III[13], John C. Geary[4], Ronald L. Gilliland[14], Michael R. Haas[1], Jennifer R. Hall[15], Matthew J. Holman[4], David G. Koch[1], David W. Latham[4], Eric Lopez[2], Sean McCauliff[15], Neil Miller[2], Robert C. Morehead[3], Elisa V. Quintana[7], Darin Ragozzine[4], Dimitar Sasselov[4], Donald R. Short[6], and Jason H. Steffen[16]

[1]NASA Ames Research Center, Moffett Field, CA, 94035, USA;  [2]UCO/Lick Observatory, University of California, Santa Cruz, CA 95064, USA;  [3]University of Florida, 211 Bryant Space Science Center, Gainesville, FL 32611-2055, USA;  [4]Harvard-Smithsonian Center for Astrophysics, 60 Garden Street, Cambridge, MA 02138, USA;  [5]Department of Astronomy, UC Berkeley, Berkeley, CA 94720, USA;  [6]San Diego State University, 5500 Campanile Drive, San Diego, CA 92182, USA;  [7]SETI Institute/NASA Ames Research Center, Moffett Field, CA 94035, USA;  [8]Department of Physics and Astronomy, San Jose State University, One Washington Square, San Jose, CA 95192, USA;  [9]Niels Bohr Institute, Copenhagen University, Juliane Maries Vej 30, DK-2100 Copenhagen, Denmark;  [10]McDonald Observatory, The University of Texas at Austin, Austin TX 78712-0259, USA;  [11]Lowell Observatory, 1400 W. Mars Hill Road, Flagstaff, AZ 86001, USA;  [12]Bay Area Environmental Research Inst./NASA Ames Research Center, Moffett Field, CA 94035, USA;  [13]Jet Propulsion Laboratory, 4800 Oak Grove Drive, Pasadena, CA 91109, USA;  [14]Space Telescope Science Institute, 3700 San Martin Drive, Baltimore, MD 21218, USA;  [15]Orbital Sciences Corporation/NASA Ames Research Center, Moffett Field, CA 94035, USA;  [16]Fermilab Center for Particle Astrophysics, MS 127, PO Box 500, Batavia, IL 60510, USA

**When an extrasolar planet passes in front of its star (transits), its radius can be measured from the decrease in starlight and its orbital period from the time between transits. Multiple planets transiting the same star reveal more: period ratios determine stability and dynamics, mutual gravitational interactions reflect planet masses and orbital shapes, and the fraction of transiting planets observed as multiples has implications for the planarity of planetary systems. But few stars have more than one known transiting planet, and none has more than three. Here we report *Kepler* spacecraft observations of a single Sun-like star that reveal six transiting planets, five with orbital periods between 10 and 47 days plus a sixth one with a longer period. The five inner planets are among the smallest whose masses and sizes have both been measured, and these measurements imply substantial envelopes of light gases. The degree of coplanarity and proximity of the planetary orbits imply energy dissipation near the end of planet formation.**

*Kepler* is a 0.95 m aperture space telescope using transit photometry to determine the frequency and characteristics of planets and planetary systems[1,2,3,4]. The only fully validated multiple transiting planet system to appear in the literature to date is Kepler-9, with two giant planets[5] orbiting exterior to a planet whose radius is only 1.6 times that of Earth[6]. The Kepler-10 system[7] contains one confirmed planet and an additional unconfirmed planetary candidate. Lightcurves of five other *Kepler* target stars, each with two or three (unverified) candidate transiting planets, have also been published[8]. A catalog of all candidate planets, including targets with multiple candidates, is presented in Borucki et al. (in preparation).

We describe below a six-planet system orbiting a star that we name Kepler-11. First, we discuss the spacecraft photometry on which the discovery is based. Second, we summarize stellar properties, primarily constrained using ground-based spectroscopy. Then, we show that slight deviations of transit times from exact periodicity due to mutual gravitational interactions confirm the planetary nature of the five inner candidates and provide mass estimates. Next, the outer planet candidate is validated by computing an upper bound on the probability that it could result from known classes of astrophysical false positives. We then assess the dynamical properties of the system, including long-term stability, eccentricities, and relative inclinations of the planets' orbital planes. We conclude with a discussion of constraints on the compositions of the planets and the clues that the compositions of these planets and their orbital dynamics provide for the structure and formation of planetary systems.

## *Kepler* Photometry

The lightcurve of the target star Kepler-11 is shown in Figure 1. After detrending, six sets of periodic dips of depth roughly 1 millimagnitude (0.1%) can be seen. When the curves are phased with these six periods, each set of dips (Figure 2) is consistent with a model[9] of a dark, circular disk masking out light from the same limb-darkened stellar disk; i.e., with the lightcurve revealing multiple planets transiting the same star. We denote the planets in order of increasing distance from the star Kepler-11b, Kepler-11c, ..., Kepler-11g.

Background eclipsing binary stars can mimic the signal of a transiting planet[10]. *Kepler* returns data for each target as an array of pixels, enabling post-processing on the ground to determine the shift, if any, of location of the target during the apparent transits. For all six planetary candidates of Kepler-11, these locations are coincident, with 3σ uncertainties of 0.7 arcsecond or less for the four largest planets and 1.4 arcseconds for the two smallest planets; see the Supplementary Information (SI) for details. This lack of displacement during transit substantially restricts the phase space available for background eclipsing binary star false positives.

Table S2 (in the SI) lists the measured transit depths and durations for each of the planets. The durations of the drops in flux caused by three of the planets are consistent with near-central transits of the same star by planets on circular orbits. Kepler-11e's transits are one-third shorter than that expected, implying an inclination to the plane of the sky of 88.8° (the eccentricity needed to explain this duration for a central transit would destabilize the system). The transit durations of planets Kepler-11b and f transits suggest somewhat non-central transits. In sum, the lightcurve shapes imply that the system is not perfectly coplanar: Kepler-11g and e are mutually inclined by at least ~0.6°.

**Ground-based Spectroscopy**

We performed a standard spectroscopic analysis[11,12] of a high resolution spectrum of Kepler-11 taken at the Keck I telescope. We derive an effective temperature, $T_{eff}$ = 5680±100 K, surface gravity, log $g$ = 4.3±0.2 (cgs), metallicity, [Fe/H] = 0.0±0.1 dex, and projected stellar equatorial rotation $v \sin i$ = 0.4±0.5 km/s. Combining these measurements with stellar evolutionary tracks[13,14] yields estimates of the star's mass, $M_\star$ = 0.95±0.10 $M_\odot$, and radius, $R_\star$ = 1.1±0.1 $R_\odot$, where the subscript $\odot$ signifies solar values. Estimates of the stellar density based upon transit observations are consistent with these spectroscopically-determined parameters. Therefore, we adopt these stellar values for the rest of the paper, and note that the planet radii scale linearly with the stellar radius. Additional details on these studies are provided in the SI.

**Transit Timing Variations**

Transits of a single planet on a Keplerian orbit about its star must be strictly periodic. In contrast, the gravitational interactions among planets in a multiple planet system cause orbits to speed up and slow down by small amounts, leading to deviations from exact periodicity of transits[15,16]. Such variations are strongest when planetary orbital periods are commensurate or nearly so, which is the case for the giant planets Kepler-9b and c[5], or when planets orbit close to one another, which is the case for the inner five transiting planets of Kepler-11.

Transit times of all six planets are listed in Table S2. Deviations of these times from the best-fitting linear ephemeris (transit timing variations, or TTVs) are plotted in Figure 3. We modeled these deviations with a system of coplanar, gravitationally-interacting planets using numerical integrations[5,17] (SI). The TTVs for each planet are dominated by the perturbations from its immediate neighbors (Figure S5). The relative periods and phases of each pair of planets, and to a lesser extent the small eccentricities, determine the shapes of the curves in Figure 3; the mass of each perturber determines the amplitudes. Thus this TTV analysis allows us to estimate the masses of the inner five planets and to place constraints on their eccentricities. We report the main results in Table 1 and detailed fitting statistics in the SI (Figure S5 and associated text).

Perturbations of planets Kepler-11d and f by planet Kepler-11e are clearly observed. These variations confirm that all three sets of transits are produced by planets orbiting the same star and yield a 4σ detection of the mass of Kepler-11e. Somewhat weaker perturbations are observed in the opposite direction, yielding a 3σ detection of the mass of Kepler-11d and a 2σ detection of the mass of Kepler-11f.

The inner pair of observed planets, Kepler-11b and c, lie near a 5:4 orbital period resonance and strongly interact with one another. The degree to which they deviate from exact resonance determines the frequency at which their TTVs should occur. Even though the precision of individual transit times is low due to small transit depths, transit-timing periodograms of both planets show peak power at the expected frequency (Figure S4). The TTVs thus confirm that Kepler-11b and c are planets, confirm that they orbit the same star, and yield 2σ determinations of their masses. The outer planet, Kepler-11g, does not strongly interact with the others; it would need to be unexpectedly massive (~ 1 $M_{Jupiter}$) to induce a detectable ($\Delta\chi^2 = 9$) signal on the entire set of transit mid-times.

**Validation of Planet Kepler-11g**

The outer planetary candidate is well-separated from the inner five in orbital period, and its dynamical interactions are not manifested in the data presently available. Thus, we only have a weak upper bound on its mass, and unlike the other five candidates, its planetary nature is not confirmed by dynamics. The signal (bottom panel of Figure 2) has the characteristics of a transiting planet and is far too large to have a non-negligible chance of being due to noise, but the possibility that it could be an astrophysical false-positive must be addressed. In order to obtain a Bayesian estimate of the probability that the events seen are due to a sixth planet transiting the star Kepler-11, we must compare estimates of the *a priori* likelihood of such a planet and of a false positive. This is the same basic methodology as was used to validate planet Kepler-9d[6].

We begin by using the BLENDER code[6] to explore the wide range of false positives that might mimic the Kepler-11g signal, by modeling the light curve directly in terms of a blend scenario. The overwhelming majority of such configurations are excluded by BLENDER. We then use all other observational constraints to further rule out blends, and we assess the *a priori* likelihood of the remaining false positives. Two classes of false positives were considered: (1) The probability of an eclipsing pair of objects that is physically-associated with Kepler-11 providing as good a fit to the Kepler data as provided by a planet transiting the primary star was found to be $0.31 \times 10^{-6}$. (2) The probability that a background eclipsing binary or star+planet pair yielding a signal of appropriate period, depth, and shape could be present and not have been detected as a result of a centroid shift in the in-transit data, or other constraints from spectroscopy and photometry, was found to be $0.58 \times 10^{-6}$. Thus the total *a priori* probability of a signal mimicking a planetary transit is $0.89 \times 10^{-6}$. There is a $0.5 \times 10^{-3}$ *a priori* probability of a transiting sixth planet in the mass-period domain. This value was conservatively estimated (not accounting for the coplanarity of the system; the value would increase by an order of magnitude if we were to assume an inclination distribution consistent with seeing transits of the five inner planets) using the observed distribution of extrasolar planets[18,19]. Details on these calculations are presented in the SI. Taking the ratio of these probabilities yields a total false alarm probability of $1.8 \times 10^{-3}$, which is small enough for us to consider Kepler-11g to be a validated planet.

**Long-Term Stability and Coplanarity**

One of the most striking features about the Kepler-11 system is how close the orbits of the planets are to one another. From suites of numerical integrations[20], dynamical survival of systems with more than three comparably-spaced planets for at least $10^{10}$ orbits has been shown only if the relative spacing between orbital semi-major axes $(a_o - a_i)/a_o$ exceeds a critical number ($\Delta_{crit} >\sim 9$) of mutual Hill-spheres $((M_i + M_o)/3M_\star)^{1/3}$, where the $a$'s and $M$'s refer to the semi-major axes and masses of the inner (i) and outer (o) planets, respectively. All of the observed pairs of planets satisfy this criterion, apart from the inner pair, Kepler-11b and c. These two planets are far enough from one another to be Hill stable in the absence of other bodies[21] (i.e., the three body problem), and they are distant enough from the other planets that interactions between the subsystems are likely to be weak. Thus stability is possible, although by no means assured. So we integrated several systems that fit the data (given in Table S4) for $2.5 \times 10^8$ years, as detailed in the SI. Weak chaos is evident both in the mean motions and the eccentricities, but the variations are at a low enough level to be consistent with long-term stability.

It is also of interest to determine whether this planetary system truly is as nearly coplanar as the Solar System, or perhaps even more so. Given that the planets all transit the star, they individually must have nearly edge-on orbits. As discussed above, the duration of planet Kepler-11e's transit implies an inclination to the plane of the sky of 88.8°, those of the two innermost planets suggest a comparable inclination, whereas those of the three other planets indicate smaller values. But even though the inclinations to the line of sight of all six planetary orbits are small, they could be rotated around the line of sight and mutually inclined to each other. The more mutually inclined a given pair of planets is, the smaller the probability that multiple planets transit[22,23]. We therefore ran Monte Carlo simulations to assess the probability of a randomly-positioned observer viewing transits of all five inner planets assuming that relative planetary inclinations were drawn from a Rayleigh distribution about a randomly selected plane. Results, displayed in Figure 4 and Table S6, suggest a mean mutual inclination of 1-2°. Details on these calculations are provided in the SI.

Mutual inclinations around the line of sight give rise to inclination changes, which would manifest themselves as transit duration changes[24]. We notice no such changes. The short baseline, small signal to noise, and small planet masses, render these dynamical constraints weak at the present time for all planets but Kepler-11e. Planet Kepler-11e has the only measured inclination, and we find that the transit duration does not change by more than 2% over the time span of the light curve. If planet Kepler-11e's orbit were rotated around the line of sight by just 2° compared to all the other components of the system, then with the masses listed in Table 1 the other planets would exert sufficient torque its orbit to violate this limit.

**Planet Compositions and Formation**

Although the Kepler-11 planetary system is extraordinary, it also tells us much about the ordinary. Measuring the both the radii and masses of small planets is extremely difficult, especially for cooler worlds farther from their star that are not heated above 1000 K. (Very high temperatures can physically alter planets, producing anomalous properties.) The planetary sizes obtained from transit depths and planetary masses from dynamical interactions together yield insight into planetary composition.

Figure 5 plots radius as a function of mass for the five newly-discovered planets whose masses have been measured. Compared to Earth, each of these planets is large for its mass. Most of the volume of each of the planets Kepler-11c-f is occupied by low-density material. It is often useful to think of three classes of planetary materials, from relatively high to low density: rocks/metals, "ices" dominated by $H_2O$, $CH_4$, and $NH_3$, and H/He gas. All of these components could have been accumulated directly from the protoplanetary disk during planet formation. Hydrogen and steam envelopes can also be the product of chemical reactions and out-gassing of rocky planets, but only up to 6% and 20% by mass, respectively[25]. In the Kepler-11 system, the largest planets with measured masses, d and e, must contain large volumes of H, as must low-mass planet f. Planets Kepler-11b and c could either be rich in "ices" (likely in the fluid state, as in Uranus and Neptune) and/or a H/He mixture. (The error bars on mass and radius for Kepler-11b allow for the possibility of an iron-depleted nearly pure silicate composition, but we view this as highly unlikely on cosmogonic grounds.) In terms of mass, all five of these planets must be primarily composed of elements heavier than helium. Future atmospheric characterization to decipher between H-dominated or steam atmospheres would tell us more about the planets' bulk composition and atmospheric stability[26].

Planets Kepler-11b and c have the largest bulk densities and would need the smallest mass fraction of hydrogen to fit their radii. Using an energy-limited escape model[27], we estimate a hydrogen mass-loss rate of several $\times 10^9$ g/s for each of the five inner planets, leading to the loss of ~0.1 $M_\oplus$, where $\oplus$ signifies the Earth, of hydrogen over 10 Gyr. This is less than a factor of 10 away from total atmosphere loss for several of the planets. The modeling of hydrogen escape for strongly irradiated exoplanets is not yet well-constrained by observations[28,29], so larger escape rates are possible. This suggests the scenario that planets Kepler-11b and c had larger H-dominated atmospheres in the past and lost these atmospheres during an earlier era when the planets had larger radii, lower bulk density, and a more active primary star, which would all favor higher mass-loss rates. The comparative planetary science allowed by the planets in Kepler-11 system may allow for advances in understanding these mass-loss processes.

The inner five observed planets of the Kepler-11 planetary system are quite densely-packed dynamically, in that significantly closer orbits would not be stable for the billions of years that the star has resided on the main sequence. The eccentricities of these planets are small, and the inclinations very small. The planets are not locked into low-order mean motion resonances.

Kepler-11 is a remarkable planetary system whose architecture and dynamics provide clues to its formation. The significant light gas component of planets Kepler-11d, e and f imply that at least these three bodies formed before the gaseous component of their protoplanetary disk dispersed, probably taking no longer than a few million years to grow to near their present masses.  The small eccentricities and inclinations of all five inner planets imply dissipation during the late stages of the formation/migration process, which means that gas and/or numerous bodies much less massive than the current planets were present.  The lack of strong orbital resonances argues against slow, convergent migration of the planets, which would lead to trapping in such configurations, although dissipative forces could have moved the inner pair of planets out from the nearby 5:4 resonance[30]. *In situ* formation would require a massive protoplanetary disk of solids near the star and/or trapping of small solid bodies whose orbits were decaying towards the star as a result of gas drag; it would also require accretion of significant amounts of gas by hot small rocky cores, which has not been demonstrated. (The temperature this close to the growing star would have been too high for ices to have condensed.)   The *Kepler* spacecraft is scheduled to continue to return data on the Kepler-11 planetary system for the remainder of its mission, and the longer temporal baseline afforded by these data will allow for more accurate measurements of the planets and their interactions.

**References**
1. Borucki, W.J. *et al*. Kepler Planet-Detection Mission: Introduction and First Results. *Science*, **327**, 977-980 (2010).
2. Koch, D.G. *et al*.  Kepler Mission Design, Realized Photometric Performance, and Early Science. *Astrophys. J.* **713**, L79-L86 (2010a).
3. Jenkins, J. *et al*. Overview of the Kepler Science Processing Pipeline. *Astrophys. J.* **713**, L87-L91 (2010).
4. Caldwell, D. *et al*.  Instrument Performance in Kepler's First Months.  *Astrophys. J.* **713**, L92-L96 (2010).
5. Holman, M.J., *et al*. Kepler-9: A System of Multiple Planets Transiting a Sun-Like, Confirmed by Timing Variations. *Science* **330**, 51-54 (2010).
6. Torres, G. *et al*.  Modeling Kepler transit light curves as false positives: Rejection of blend scenarios for Kepler-9, and validation of Kepler-9d, a super-Earth-size planet in a multiple system. *Astrophys. J.* in press, arXiv:1008.4393 (2011).
7. Batalha, N. *et al*. Kepler's First Rocky Planet: Kepler-10b.  *Astrophys. J.* in press, (2011) .
8. Steffen, J.H. *et al*. Five Kepler Target Stars That Show Multiple Transiting Exoplanet Candidates. *Astrophys. J.* **725**, 1226-1241 (2010).
9. Mandel, K. & Agol, E.  Analytic Light Curves for Planetary Transit Searches. *Astrophys. J.* **580**, L171-L175 (2002).
10. Brown, T.  Expected Detection and False Alarm Rates for Transiting Jovian Planets.  *Astrophys. J.* **593**, L125-L128 (2003).
11. Valenti, J.A. & Piskunov, N. Spectroscopy made easy: A new tool for fitting observations with synthetic spectra. *Astron. Astrophys. Supp.* **118**, 595-603 (1996).
12. Valenti, J.A. & Fischer, D.A. Spectroscopic Properties of Cool Stars (SPOCS). I. 1040 F, G, and K Dwarfs from Keck, Lick, and AAT Planet Search Programs. *Astrophys. J. Supp.* **159**, 141-166 (2005).

13. Girardi, L., Bressan, A., Bertelli, G., Chiosi, C. Evolutionary tracks and isochrones for low- and intermediate-mass stars: From 0.15 to 7 $M_{sun}$, and from Z=0.0004 to 0.03, *Astron. Astrophys. Supp.* **141** 371-383 (2000).
14. Yi, S. *et al.* Toward Better Age Estimates for Stellar Populations: The $Y^2$ Isochrones for Solar Mixture. *Astrophys. J. Supp.* **136**, 417-437 (2001).
15. Holman, M.J. & Murray, N.W. The Use of Transit Timing to Detect Terrestrial-Mass Extrasolar Planets. *Science*, **307**, 1288-1291 (2005).
16. Agol E., Steffen, J., Sari, R., Clarkson, W. On detecting terrestrial planets with timing of giant planet transits, *Mon. Not. R. Astron. Soc.* **359**, 567-579 (2005).
17. Fabrycky, D. Non-Keplerian Dynamics, *EXOPLANETS*, ed. S. Seager, University of Arizona Press, 217-238 (2010).
18. Cumming, A. *et al.* The Keck Planet Search: Detectability and the Minimum Mass and Orbital Period Distribution of Extrasolar Planets. *Pub. Astron. Soc. Pacific* **120**, 531-554 (2008).
19. Howard, A. *et al.* The Occurrence and Mass Distribution of Close-in Super-Earths, Neptunes, and Jupiters. *Science*, **330**, 653-655 (2010).
20. Smith, A.W. & Lissauer, J.J. Orbital stability of systems of closely-spaced planets. *Icarus* **201**, 381-394 (2009).
21. Gladman, B. Dynamics of systems of two close planets. *Icarus* **106**, 247-2 (1993).
22. Koch, D. and Borucki, W. A Search For Earth-Sized Planets In Habitable Zones Using Photometry, *First International Conf on Circumstellar Habitable Zones,* Travis House Pub., 229, (1996).
23. Ragozzine D. & Holman, M. J. The Value of Systems with Multiple Transiting Planets. sumitted to *Astrophys. J.*, arXiv:1006.3727 (2010).
24. Miralda-Escude, J. Orbital Perturbations of Transiting Planets: A Possible Method to Measure Stellar Quadrupoles and to Detect Earth-Mass Planets. *Astrophys. J.* **564**, 1019-1023 (2002).
25. Elkins-Tanton, L. T. & Seager, S. Ranges of Atmospheric Mass and Composition of Super-Earth Exoplanets. *Astrophys. J.* **685**, 1237-1246 (2008).
26. Miller-Ricci, E., Seager, S., Sasselov, D. The Atmospheric Signatures of Super-Earths: How to Distinguish Between Hydrogen-Rich and Hydrogen-Poor Atmospheres. *Astrophys. J.* **690**, 1056-1067 (2009).
27. Lecavelier Des Etangs, A. A diagram to determine the evaporation status of extrasolar planets. *Astron. Astrophys.* **461**, 1185-1193 (2007).
28. Vidal-Madjar, A., et al. An extended upper atmosphere around the extrasolar planet HD209458b, *Nature*, **422**, 143-146 (2003).
29. Lecavelier Des Etangs, A., *et al.* Evaporation of the planet HD 189733 observed in H I Lyman-α. *Astron. Astrophys.* **514**, id.A72 (2010).
30. Papaloizou, J. C. B. & Terquem, C. On the dynamics of multiple systems of hot super-Earths and Neptunes: tidal circularization, resonance and the HD 40307 system. *Mon. Not. R. Astron. Soc.*, **405**, 573-592 (2010).
31. Rowe, J. F. *et al.* Kepler Observations of Transiting Hot Compact Objects. *Astrophys. J.* **713**, L150-L154 (2010).
32. Miller, N., Fortney, J. J. & Jackson, B. Inflating and Deflating Hot Jupiters: Coupled Tidal and Thermal Evolution of Known Transiting Planets. *Astrophys. J.* **702**, 1413–1427 (2009).

33. Nettelmann, N., Fortney, J. J., Kramm, U., Redmer, R. Thermal evolution and interior models of the transiting super-Earth GJ 1214b. arXiv:1010.0277, submitted to *Astrophys. J.* (2010).
34. Rogers, L. A. & Seager, S. A Framework for Quantifying the Degeneracies of Exoplanet Interior Compositions. *Astrophys. J.* **712**, 974-991 (2010).

**Supplementary Information** is linked to the online version of the paper at www.nature.com/nature.

**Acknowledgements**
*Kepler* was competitively selected as the tenth Discovery mission. Funding for this mission is provided by NASA's Science Mission Directorate. The authors thank the many people who gave so generously of their time to make the *Kepler* mission a success. A. Dobrovolskis, T.J. Lee, and D. Queloz provided constructive comments on the manuscript. D. C. F. and J. A. C. acknowledge NASA support through Hubble Fellowship grants #HF-51272.01-A and #HF-51267.01-A, respectively, awarded by STScI, operated by AURA under contract NAS 5-26555.

**Author Information**

J. Lissauer, jack.lissauer@nasa.gov

D. Fabrycky, fabrycky@ucolick.org

**Author Contributions**

**J. Lissauer** led research effort to confirm/validate candidates as planets. Assisted in dynamical study. Wrote plurality of manuscript. Developed and wrote up majority of interpretation. **D. Fabrycky** performed dynamical analysis on transit times and derived planetary masses, derived dynamical constraint on mutual inclinations. Performed long-term stability calculations. Designed and made figures 1-3, wrote significant portion of the SI, including dynamics figures. **E. Ford** measured transit times, including special processing for Q6 data, checked for transit duration variations, contributed to interpretation, supervised transit probability and relative inclination analysis. The following 7 authors contributed equally: **W. Borucki** developed photometers, observational techniques, and analysis techniques that proved *Kepler* could succeed. Participated in the design, development, testing and commissioning of the *Kepler* mission and participated in the evaluation of the candidates that led to the discovery of this system. **F. Fressin** modeled *Kepler* transit light curves as false positives leading to rejection of blend scenarios for hierarchical triple and background configurations. **G. Marcy** obtained and reduced spectra that yield the properties of the star. **J. Orosz** measured planet radii and impact parameters. **J. Rowe** performed transit search to identify candidates. Multi-candidate lightcurve modeling to determine stellar and planetary parameters. Transit timing measurements. **G. Torres** modeled *Kepler* transit light curves as false positives leading to rejection of blend scenarios for hierarchical triple and background configurations. **W. Welsh** measured transit times and O-C curves and used Monte Carlo to determine robust uncertainties. The remaining authors listed below contributed equally: **N. Batalha** directed target selection, KOI inspection, tracking, and vetting. **S. Bryson** supported centroid and light curve analysis and participated in validation discussions. **L. Buchhave** took and analyzed first reconnaissance spectrum of target star. **D. Caldwell** worked on definition and development of the Science Operations Center analysis pipeline. **J. Carter** assisted in the determination of transit times and durations from the Kepler photometry. **D. Charbonneau** provided advice on blender analysis. **J. Christiansen** supported the science operations to collect the Kepler data. **W. Cochran** obtained, reduced and analyzed reconnaissance spectroscopy. **J. Desert** participated in blend studies. **E. Dunham** provided optical, electronic, and systems support for flight segment, commissioning work, and discussions regarding validation of small planets. **M. Fanelli** reviewed light curves and centroid time series and participated in verification and validation of the science pipeline. **J. Fortney** modeled the interior structure and mass-radius relations of the planets. Wrote text on interpreting planetary structure. **T. Gautier III** performed follow-up observation support, commissioning work. **J. Geary** worked on the design of the *Kepler* focal plane and associated CCD imagers and electronics. **R. Gilliland** performed difference image based centroid analysis as means of discriminating against background eclipsing binary stars. **M. Haas** led the team responsible for the scientific commissioning and operation of the instrument, and processing the data to produce light curves. **J. Hall** developed operations procedures and performed processing of *Kepler* data to produce light curves. **M. Holman** developed trending algorithm; helped assembling and writing results. **D. Koch** designed and developed major portions of the *Kepler* mission. **D. Latham** led reconnaissance spectroscopy, stellar classification; preparation of the Kepler Input Catalog. **E. Lopez** modeled the interior structure and mass-radius relations of the planets. **S. McCauliff** wrote software to manage and archive pixel and flux time series data. **N. Miller** modeled the interior structure and mass-radius relations of the planets. **R. Morehead** performed transit probability vs. relative inclination analysis. **E. Quintana** wrote software to calibrate the pixel data to generate the flux time series. **D. Ragozzine** conducted data analysis and interpretation. **D. Sasselov** performed calculations using stellar evolution models to determine stellar parameters. **D. Short** developed and used code to measure transit times. **J. Steffen** worked on validation of analysis methods and composition of text. The authors declare no competing financial interests.

**Table**

Table 1 Planet Properties

| Planet | Period (days) | Epoch (BJD) | Semi-major Axis (AU) | Inclination (degrees) | Transit Duration (hrs) | Transit Depth (mmag) | Radius (R$_\oplus$) | Mass (M$_\oplus$) | Density (g/cm³) |
|---|---|---|---|---|---|---|---|---|---|
| b | 10.30375 ± 0.00016 | 2454971.5052 ± 0.0077 | 0.091 ± 0.003 | 88.5 +1.0,-0.6 | 4.02 ± 0.08 | 0.31 ± 0.01 | 1.97 ± 0.19 | 4.3 +2.2,-2.0 | 3.1 +2.1,-1.5 |
| c | 13.02502 ± 0.00008 | 2454971.1748 ± 0.0031 | 0.106 ± 0.004 | 89.0 +1.0,-0.6 | 4.62 ± 0.04 | 0.82 ± 0.01 | 3.15 ± 0.30 | 13.5 +4.8,-6.1 | 2.3 +1.3,-1.1 |
| d | 22.68719 ± 0.00021 | 2454981.4550 ± 0.0044 | 0.159 ± 0.005 | 89.3 +0.6,-0.4 | 5.58 ± 0.06 | 0.80 ± 0.02 | 3.43 ± 0.32 | 6.1 +3.1,-1.7 | 0.9 +0.5,-0.3 |
| e | 31.99590 ± 0.00028 | 2454987.1590 ± 0.0037 | 0.194 ± 0.007 | 88.8 +0.2,-0.2 | 4.33 ± 0.07 | 1.40 ± 0.02 | 4.52 ± 0.43 | 8.4 +2.5,-1.9 | 0.5 +0.2,-0.2 |
| f | 46.68876 ± 0.00074 | 2454964.6487 ± 0.0059 | 0.250 ± 0.009 | 89.4 +0.3,-0.2 | 6.54 ± 0.14 | 0.55 ± 0.02 | 2.61 ± 0.25 | 2.3 +2.2,-1.2 | 0.7 +0.7,-0.4 |
| g | 118.37774 ± 0.00112 | 2455120.2901 ± 0.0022 | 0.462 ± 0.016 | 89.8 +0.2,-0.2 | 9.60 ± 0.13 | 1.15 ± 0.03 | 3.66 ± 0.35 | < 300 | - |

Planetary periods and transit epochs are the best-fitting linear ephemerides. Uncertainty in epoch is median absolute deviation of transit times from this ephemeris; uncertainty in period is this quantity divided by number of orbits between first and last observed transits. Radii are from Table S2; uncertainties represent 1σ ranges, and are dominated by uncertainties in the radius of the star. The mass estimates are the uncertainty-weighted means of the three dynamical fits (Table S4) to TTV observations (Table S2) and the quoted ranges cover the union of 1σ ranges of these three fits. One of these fits constrains all of the planets to be on circular orbits, the second one allows only planets Kepler-11b and c to have eccentric orbits, and the third solves for the eccentricities of all five planets b-f; see SI. Stability considerations may preclude masses near the upper ends of the quoted ranges for the closely-spaced inner pair of planets. Inclinations are with respect to the plane of the sky.

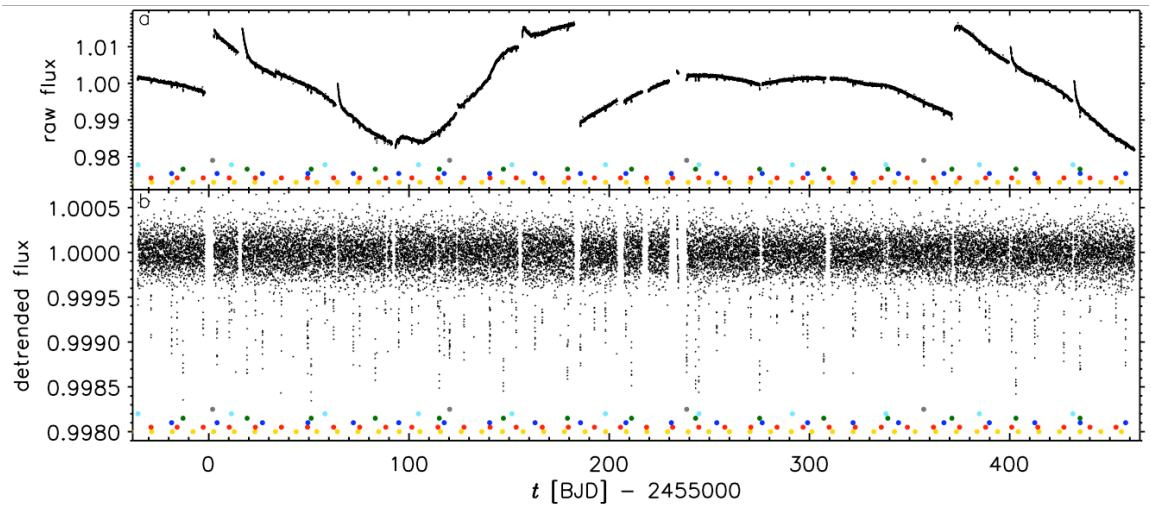

Figure 1:
Lightcurves of Kepler-11, raw and detrended. Kepler-11 is a G dwarf star with Kepler magnitude $Kp$ = 13.7, visual magnitude V = 14.2 mag, and celestial coordinates RA = $19^h 48^m 27.62^s$, Dec = +41° 54' 32.9"; alternate designations used in catalogs are KIC 6541920 and KOI-157. Kepler-11 is ~ 2000 light-years from Earth. Variations in the brightness of Kepler-11 have been monitored with an effective duty cycle of 91% over the time interval barycentric Julian date (BJD) 2454964.511- 2455462.296, with data returned to Earth at a cadence of 29.426 minutes (long cadence, LC). Shown above are *Kepler* photometric data in 30-minute intervals, raw from the spacecraft with each quarter normalized to its median (*top*) and after detrending with a polynomial filter[31] (Rowe *et al.* 2010). These data are available from the MAST archive at http://archive.stsci.edu/kepler/ . Note the difference in vertical scales between the two panels. The six sets of periodic transits are indicated with dots of differing colors. Four photometric datapoints representing the triply concurrent transit of planets Kepler-11b, d and e at BJD 2455435.2 (Figure S6) are not shown as their values lie below the plotted range. Data have also been returned for this target star at a cadence of 58.85 seconds since BJD 2455093.215, but our analysis is based exclusively on the LC data.

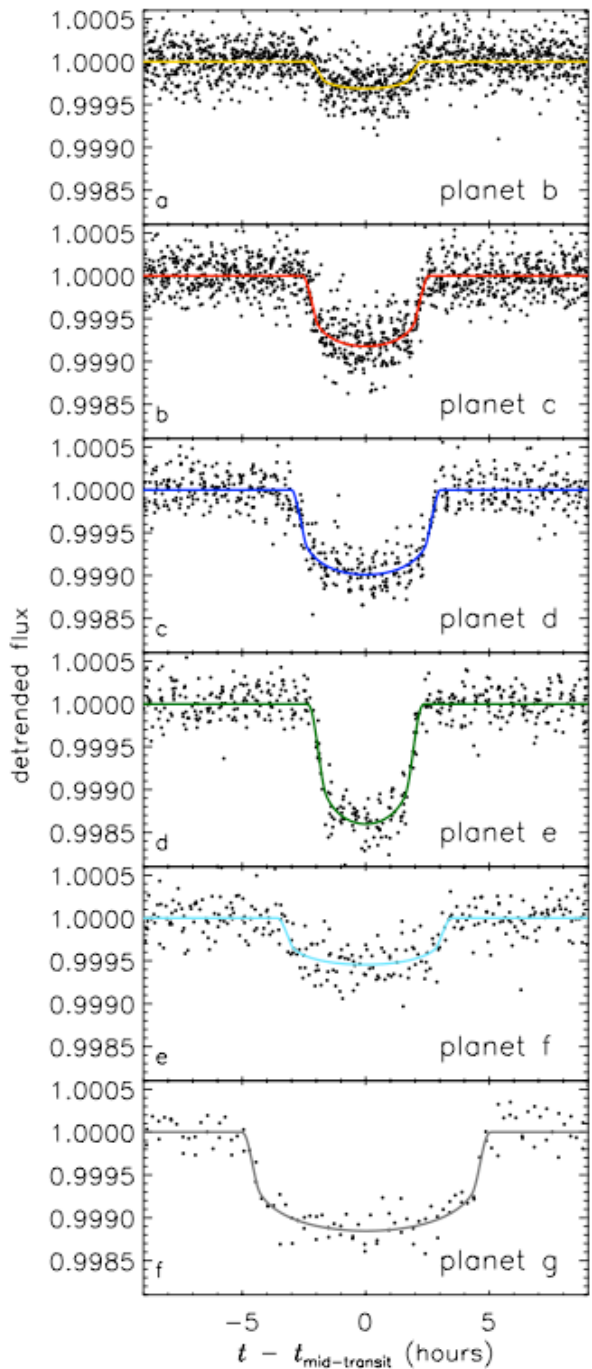

Figure 2:
Detrended data of Figure 1 shown phased at the period of each transit signal and zoomed to an 18-hour region around mid-transit. Overlapping transits are not shown, nor were they used in the model. Each panel has an identical vertical scale, to show the relative depths, and identical horizontal scale, to show the relative durations. The color of each planet's model lightcurve matches the color of the dots marking its transits in Figure 1.

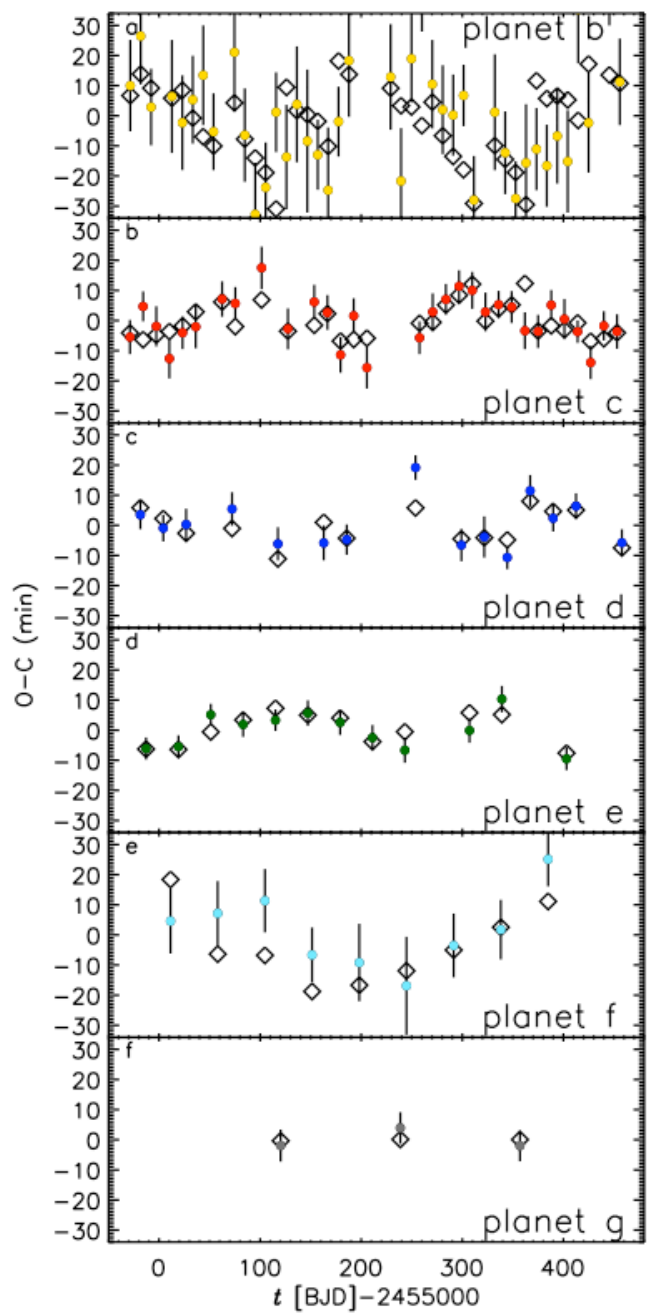

Figure 3:
Transit timing variations and dynamical fits. Observed (O) mid-times of planetary transits (see the SI for transit-fitting method) minus a Calculated (C) linear ephemeris, are plotted as dots with 1σ error bars; colors correspond to the planetary transit signals in Figures 1 and 2. The times derived from the "circular fit" model described in Table S4 are given by the open diamonds. Contributions of individual planets to these variations are shown in Figure S5.

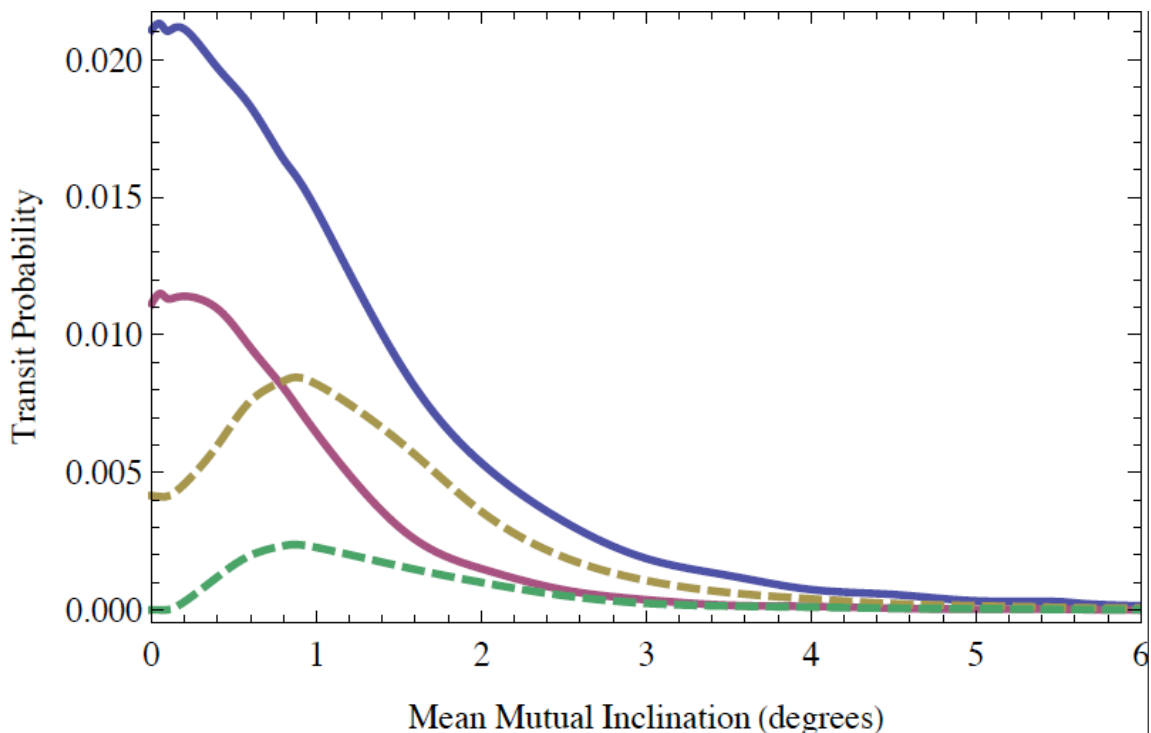

Figure 4:
Transit probabilities as a function of relative orbital inclinations of planets orbiting Kepler-11. Results of Monte Carlo simulations to assess the probability of a randomly-positioned observer viewing transits of various combinations of observed and hypothesized planets around the star Kepler-11, assuming that relative planetary inclinations were drawn from a Rayleigh distribution about a randomly selected plane. The solid blue curve shows probabilities the five inner planets (Kepler-11b-f) transiting. The solid pink curve shows probabilities all six inner planets to be seen transiting. The ratio of the orbital period of planet Kepler-11g to that of Kepler-11f is substantially larger than that for any other neighboring pair of transiting planets in the system. If we hypothesize that a seventh planet orbits between these objects with a period equal to the geometric mean of planets Kepler-11f and g, then the probability of observing transits of any combination totaling six of these seven planets is shown in the dashed golden curve. The dashed green curve shows the probability for the specific observed six to transit. Details on these calculations are provided in the SI.

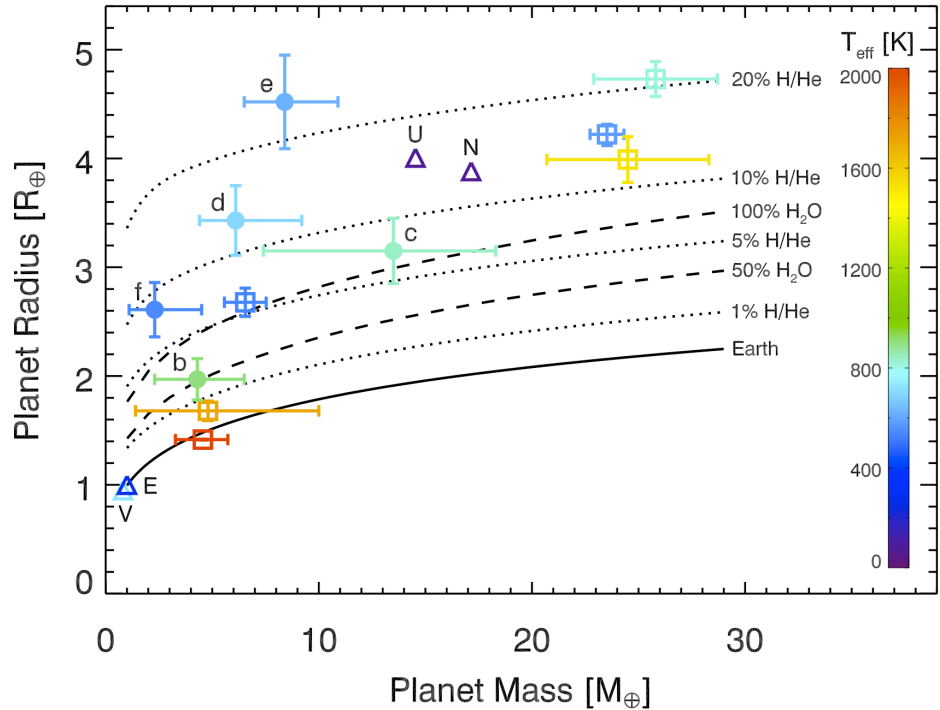

Figure 5:
Mass-radius relationship of small transiting planets, with Solar System planets shown for comparison. Planets Kepler-11b-f are represented by filled circles with 1σ error bars, with letters written above; values and ranges are as given in Table 1. Other transiting extrasolar planets in this size range are shown as open squares, representing in order of ascending radius Kepler-10b, CoRoT-7b, GJ 1214b, Kepler-4b, GJ 436b, and HAT-P-11b. The triangles (labeled V, E, U and N) correspond to Venus, Earth, Neptune and Uranus, respectively. The colors of the points show planetary temperatures (measured for planets in our Solar System, computed mean planet-wide equilibrium temperatures for Bond albedo = 0.2 for the extrasolar planets), with values shown in the color bar at the right. Using previously implemented planetary structure and evolution models[32,33], we plot mass-radius curves for 8 Gyr-old planets, assuming $T_{eff}$ = 700 K. The solid black curve corresponds to models of planets with Earth-like rock-iron composition. The higher dashed curve corresponds to 100% $H_2O$. All other curves use a water or $H_2$/He envelope atop the rock-iron core. The lower dashed curve is 50% $H_2O$ by mass, while the dotted curves are $H_2$/He envelopes that make up 2%, 6%, 10%, and 20% of the total mass. There is significant degeneracy in composition from only a mass and radius measurement[34]. Planet Kepler-11d, e, and f appear to require a $H_2$/He envelope, much like Uranus and Neptune, while Kepler-11b and c may have $H_2O$ and/or $H_2$/He envelopes. We note that multi-component and mixed compositions (not shown above), including rock/iron, $H_2O$, and $H_2$/He, are expected and lead to even greater degeneracy in determining composition from mass and radius alone.

# Supplemental Information

## 1. *Kepler* Photometry and Centroid Analysis

Since background eclipsing binaries (BGEB) are a common false positive for transiting planet searches, we conducted a careful analysis of the target star centroid both in and out of transit. This analysis is a useful tool to reject BGEBs since they tend to cause a significant displacement of the centroid during the transit events. Using the transit ephemeris for each planet, non-overlapping transits were selected from the Quarter 3 data (BJD 2,455,093.215–2,455,182.495). For each planet a direct image composed of exposures within its transit(s) was formed, as well as a control direct image with symmetrically chosen points outside of transit with widths equal to the transit on each side. A difference image was formed by subtracting the in-transit image from the out-of-transit control. A mean direct image was formed from the in- and out-of-transit direct images.

Pixel response function (PRF)[35] fits were then performed separately to the direct and difference images for each planet, with the resulting radial offset tabulated. Errors were established via a Monte Carlo study by using the PRF to generate a large number of synthetic direct and difference image realizations. The transit strength on each simulated transit source was set by the requirement that the resulting transit depth in the simulated flux time series match the observed transit depth. The offset of each trial transit source from the target position and the resulting centroid errors from the known target location were tabulated, producing a distribution of the centroid error vs. transit offset. The $3\sigma$ point in this distribution was then chosen to represent the circle of confusion in which false positives cannot be found. This was augmented by a 0.12 arcsec $1\sigma$ error to account for systematics. The resulting table of offsets, and corresponding significance, is shown in Table S1. The centroids are consistent for all planets with the true source of the transit being spatially coincident with the target star (none of the deviations exceeds $2\sigma$). Moreover, BGEBs beyond a radius of 1.4 arc seconds are excluded at $3\sigma$ for the two smallest planets, Kepler-11b and f, and beyond a radius of only 0.7 arc seconds for all of the other planets.

Contamination of the flux light curves (the ratio of flux in the photometric aperture from all sources excluding the target star to the flux from all sources) is computed[36] using synthetic images based on the PRF model and the *Kepler* Input Catalog (KIC[37]). The contamination estimate is approximate, with uncertainties mainly due to inaccuracies of the PSF model and catalog errors (primarily the omission of dim stars). The contamination value varies from quarter to quarter because the photometric aperture varies from quarter to quarter. When using stars listed in the KIC, the contamination is as large as 7.2% of the total flux. If no stars are missing from the KIC, then the contamination error uncertainties are less than 1% of the target star flux. The KIC is complete only to about $Kp = 17$, however. Observations from the United Kingdom Infrared Telescope (UKIRT) indicate two stars which are in or near the photometric aperture that are not in the KIC, with $Kp = 18.83 \pm 0.64$ and $Kp = 19.53 \pm 0.79$. A conservative estimate, which takes the bright value of these star's magnitude uncertainties and slightly over-estimates the contributions of these star's flux to the photometric aperture, increases the largest contamination value from 7.2% to 9.1% of the total flux. In other quarters, which have lower contamination values, these two stars are farther from the photometric aperture and thus increase the contamination by smaller amounts. Therefore we expect that stars missing from the KIC increase the contamination by no more than 1% to 2% of the total flux. The contamination value of 12.3% given in the MAST archive (in the *Kepler* Target Search summary) is larger than the values we find here, and can be taken as a firm upper limit.

## 2. Properties of the Star Kepler-11

An accurate assessment of the stellar radius and its associated uncertainty is essential for constraining the nature of the planets. We performed a standard LTE spectroscopic analysis of a high-resolution template spectrum of Kepler-11 from Keck/HIRES using the SME package ("Spectroscopy Made Easy"[11,12]) and derived an effective temperature $T_{\text{eff}} = 5680 \pm 100$ K, a surface gravity $\log g = 4.3 \pm 0.2$ (cgs), a metallicity [Fe/H] $= 0.00 \pm 0.10$ dex, and a projected equatorial rotational velocity of $v \sin i = 0.4 \pm 0.5$ km s$^{-1}$. A comparison of these parameters with stellar evolution models[14,13] yields an estimate of the star's mass, $M_\star = 0.95 \pm 0.10 \, M_\odot$, and its radius, $R_\star = 1.10 \pm 0.10 \, R_\odot$. The stellar isochrones in this part of the HR diagram are not intersecting, so they provide unique solutions. According to these models, the star appears slightly evolved and relatively old (6–10 Gyr).

The above radius, however, is essentially set by $\log g$, which is a notoriously difficult quantity to measure accurately. Since we do not have a parallax (and hence a luminosity) or asteroseismic constraint on the mean stellar density, this raises concerns about correlations with $T_{\text{eff}}$ and [Fe/H], as well as possible systematic errors. Fortunately, we are able to place bounds on the stellar radius independent of $\log g$. The transit duration is inversely proportional to the stellar density. Although the impact parameters are poorly constrained, the maximum duration occurs for a central transit and a low stellar density. Thus, the transit durations provide an independent lower limit on the stellar radius that is similar to the $1\sigma$ lower limit based on the spectroscopy/isochrone analysis. The main sequence provides yet a third independent lower limit for the stellar radius, which is comparable to the $2\sigma$ lower limit for the radius based on the spectroscopy/isochrone analysis. While our current estimate of the stellar radius is not particularly precise, using multiple independent methods allows us to gain confidence in its accuracy.

We obtained five spectra with the HIRES spectrometer on the Keck 1 telescope with a resolving power of $R = 50{,}000$ and a signal-to-noise ratio of 30, with exposures taking 20 min. A simple Doppler analysis made use of the telluric A and B bands to set the wavelength calibration zero-point, and a cross correlation was done in four spectral segments from 700–800 nm, avoiding telluric lines. Figure S1 shows the resulting velocity measurements. The mean velocity of $-57.16$ km s$^{-1}$ is measured relative to the Solar System barycenter; compared to the local velocity dispersion of $\sim 30$ km s$^{-1}$, this Galactic velocity suggests that Kepler-11 is a member of the thick disk of the Milky Way. The standard deviation of $0.34\,\mathrm{km\,s^{-1}}$ is consistent with fluctuations from the internal errors of $0.28\,\mathrm{km\,s^{-1}}$. The visually apparent downward trend is not statistically significant, but more spectra over a year time scale would be helpful to continue hunting for any sign of another star located within 1 arcsec. Clearly these constant RVs rule out the presence of another star within 2 magnitudes of Kepler-11 orbiting within 1 AU, as such a companion would cause the composite velocity to change by more than $1\,\mathrm{km\,s^{-1}}$. To be sensitive to the planets themselves, a much longer integrations or a bigger telescope would be needed: the largest $K$-amplitude we anticipate is 5.3 m s$^{-1}$, which comes from the one-$\sigma$ upper bound on the mass of planet Kepler-11c from Table 1 of the main text.

With six prospective planets, Kepler-11 raises the possibility of multiple stars within the *Kepler* pixel, with transits (or eclipses) occurring for each of those stars. We have searched for additional, unresolved stars near Kepler-11 by cross-correlating the Keck-HIRES spectrum against the solar spectrum, looking for multiple peaks in the cross-correlation function (CCF). We found no other peaks. The CCF, shown in Figure S2, exhibits a single narrow peak. The small ripples in the wings of the CCF are due to stochastic overlaps of spectral lines as one spectrum is Doppler shifted past the other. We simulated a stellar companion by constructing mock spectra of Kepler-11 as a blend of two stars by adding to the Kepler-11 spectrum an additional G-type spectrum having a fraction of the intensity of, and a Doppler shift relative to, Kepler-11. Figure S3 shows that we would have detected the spectral signature of such a star if it had a separation of $30\,\mathrm{km\,s^{-1}}$ (~1 AU) and was no more than 3 magnitudes fainter than Kepler-11 ($3\sigma$). Doppler separations of less than $20\,\mathrm{km\,s^{-1}}$ would be difficult to detect as the spectral lines overlap too much.

## 3. Transit Fits, Times, and Durations

### 3.1. Transit Times

We fit a standard transit model[9] to the light curve for each planet to measure the transit time, planet-star radius ratio, transit duration, and impact parameter, as well as the flux normalization and a local linear slope. We numerically average the model over the 30 minute integration duration. First, we fit a single model to each of the transits of each planet individually, assuming a constant orbital period. Second, we hold the radius ratio, transit duration, and impact parameter fixed, and fit a small segment of the lightcurve around each transit for the remaining parameters. We exclude data points with concurrent transits (i.e., when more than one planet transited at the same time). Third, we align the light curve using each measured transit time, and refit for the transit parameters (aside from period and epoch). We iterate the second and third steps to converge on a model.

We adopt a 4$^{\mathrm{th}}$ order non-linear limb darkening law[9] and hold the limb darkening parameters fixed based on the spectroscopic parameters. The impact parameter affects the duration of ingress and egress relative to the overall transit duration. While most of the transit parameters are well-measured, the impact parameter is only weakly constrained with the present data, due to the combination of long cadence integration time, limb darkening, and noise. Once further *Kepler* data and/or transit follow-up observations at longer wavelengths provide precise measurement of the impact parameters, it will be possible to measure the orbital inclinations, as well as to increase precision of stellar and planetary parameters (e.g., radius, density).

The resulting linear ephemeris and transit times are listed in Table S2 (a–g). We omitted times when there are overlapping transits, because we were unable to measure their times accurately. Furthermore, because of the small signal from planet Kepler-11b, we noticed problems with five transits and omitted their times. For four of these ($N = 8, 20, 26, 27$) the $\chi^2$-surface for that transit time was clearly bimodal and the results depended on smoothing length; a final transit ($N = 41$) was the biggest outlier from a linear ephemeris and had an $O - C$ value 80 minutes above the transits adjacent to it. Lastly, data gaps also caused loss of some transits.

### 3.2. Mean Profiles, Transit Durations and Depths

Mean transit profiles for each planet were made by phasing the individual transits using the transit times and mean periods determined above. Overlapping transits from other planets were masked out in this process. The transit profiles were modeled using the ELC code[38] and its various optimizers, in particular its genetic code[39], its Monte Carlo Markov chain code[40], and its "grid search" routine. Given the low masses and long periods, the star can be treated as a limb-darkened circular disk and the planets as opaque circular disks, so ELC's "analytic" mode[41], was used. For each planet, the free parameters are the inclination $i$, the ratio of radii $R_p/R_\star$, the stellar mass $M_\star$, the stellar radius $R_\star$, and a small phase shift to account for uncertainties in the ephemerides. A quadratic limb darkening law of the form

$$I(\mu)/I_0 = 1 - a(1 - \mu) - b(1 - \mu)^2 \tag{1}$$

was used, in which $\mu = \cos\theta$ (where $\theta$ is the angle from the center of the stellar disk), $I(\mu)$ is the specific intensity at the angle $\mu$, and $I_0$ is the specific intensity normal to a surface element. Initial fits showed that there was no sensitivity to the values of the coefficients. We therefore adopted the coefficients interpolated from the tables computed by Prša[42], namely $a = 0.495$ and $b = 0.178$. The models were computed at intervals of one minute, then binned to 30 minutes to mimic the sampling of the *Kepler* data.

The normalized light curves were corrected for the contamination fraction $k$ using the equation

$$F_{\text{new}}(t) = \frac{F_{\text{old}}(t) - k}{1 - k}. \tag{2}$$

We performed fits using three scenarios for the contamination:

(i) quarter-by-quarter contamination values discussed above, where

$$k_{Q1} = 0.0662,$$
$$k_{Q2} = 0.0290,$$
$$k_{Q3} = 0.0318,$$
$$k_{Q4} = 0.0722,$$
$$k_{Q5} = 0.0656,$$
$$k_{Q6} = 0.0290;$$

(ii) no contamination (e.g., $k = 0$ for all quarters); and (iii) $k = 0.10$ for all quarters. For each case, the various optimizers were run to find the minimum $\chi^2$. Once that was found, the $\chi^2$/d.o.f. of each planet ranges from 1.08–1.17, indicating that the measurement uncertainties are unlikely to have been underestimated by 4–8%. The uncertainties on the individual photometric measurements were scaled to give $\chi^2_{\rm min}$/d.o.f. $\approx 1$. The optimizers were run again, accumulating more than 50,000 models with heavy sampling in the parameter space near the $\chi^2$ minumum. Based on the spectroscopic analysis discussed in Section 2, we adopt a stellar mass of $M_\star = 0.95 \pm 0.10\,{\rm M}_\odot$ and a stellar radius of $R_\star = 1.10 \pm 0.10\,{\rm R}_\odot$. Since the optimizer codes generally work better when a uniform distribution of parameters is initially adopted, the adopted stellar mass and radius and their uncertainties were folded into the process by adding additional terms to the $\chi^2$:

$$\chi^2_{\rm total} = \chi^2_{\rm phot} + \left(\frac{M_\star/{\rm M}_\odot - 0.95}{0.10}\right)^2 + \left(\frac{R_\star/{\rm R}_\odot - 1.10}{0.10}\right)^2. \qquad (3)$$

The $1\sigma$ confidence limits on the fitting parameters and other derived parameters were computed by collapsing the $\chi^2$ surface on each parameter of interest and finding the range where $\chi^2 \leq \chi^2_{\rm min} + 1$.

Table S3 gives the results of the model fits. We give the transit depths and durations (defined as the time between first and fourth contacts), the radius ratios, the impact parameters, where $b = a/R_\star \cos i$ and $a$ is the orbital separation, the derived planetary radii, and the median photometric uncertainty after the scaling. The top section gives the parameters using the quarter-by-quarter contamination values (our adopted model). The middle and bottom sections give the results when no contamination is used and when 10% contamination is used, respectively. As expected, when no contamination is used, the true transit depths are shallower, and when the maximum plausible contamination is used, the true transit depths are deeper. Likewise, when no contamination is used, the radius ratios are smaller compared to when the maximal contamination is used, where the change is generally less than $\approx 2\sigma$ for each planet. The planetary radii in physical units (e.g., Earth radii where ${\rm R}_\oplus = {\rm R}_\odot/109.1$) depend on the adopted stellar parameters, and in this case the uncertainties on the planetary radii are dominated by the uncertainty in the stellar radius. Thus over the entire range of contamination values considered, the planetary radii change by less then $1\sigma$.

Planet Kepler-11e is the only planet whose observations enabled us to tightly limit the impact parameter: $b_e = 0.79^{+0.04}_{-0.05}$. The Kepler-11b & f solutions mildly prefer a non-zero impact parameter, but are consistent with zero, as are planets Kepler-11c, d, and g — see Figure S4.

## 4. Dynamical Confirmation of the Inner Five Planets

Here we describe the transit timing data and dynamical fits in more detail, showing how aspects of the signals led us to the conclusions of the main text, in particular that the transiting objects have planetary masses. First, the transits are not strictly periodic. That is, constant orbital periods are not sufficient to explain the timing data. Using the 106 transit times of Table S2 (b through g), and fitting $P$ and $T_0$ for each of the six transit signals (Table S2a), we find $\chi^2 = 191.51$ for 94 degrees of freedom (106 data points − 6 fitted periods − 6 fitted epochs). In principle, the error bars on individual transit times may be underestimated, accounting for this. Inflating the errors as suggested by the photometric fits (maximum 8%; Section 3.2) would give $\chi^2 \approx 164$ for the constant-period model, which is still formally unacceptable. In the following, we take the error bars of Table S2a at face value and interpret the excess variation in the transit times as a dynamical signal.

Owing to their short orbital periods, planets Kepler-11b and c have the most transits in our data set. But their small radii render the uncertainties on each transit time large, so that their transit timing variations are not obvious by eye (Figure 3, main text). From sample integrations, as well as analytic theory[16], we expect their timing curves to be dominated by their proximity to the 4:5 mean motion resonance. In particular, the frequency of the oscillations should be:

$$f_{O-C} = 4/P_b - 5/P_c = 1/(231 \text{ days}). \qquad (4)$$

We plot the periodograms[43] of the $O - C$ data for these two planets in Figure S5. The peak frequency for each of these planets is at this expected frequency, which we interpret as robust and conclusive evidence for their interaction and motivates more detailed dynamical modeling.

From this example, we find that the $O - C$ signals of the inner two transiting planets have the same shape, but are of opposite sign and have different amplitude[5]. In the case of the Kepler-11b/c pair, the period of those signals is a straightforward function of their transit periods. Similarly, we may expect that the shapes of the $O - C$ signal for all the planets depend chiefly on their periods. Therefore we numerically integrated the orbits of the planets on circular orbits, to find the functional form of the $O - C$ that they induce on each other. The signal of each transiting planet is expected to be linearly proportional to the mass of the perturbing planet[5,16,44], so we began by fitting a linear combination of these signals to the data. But the proximity of the Kepler-11b/c pair to the 4:5 resonance caused moderate non-linearity. Therefore, to fit the transit times, we used the Levenberg-Marquardt non-linear minimization algorithm to drive 6-planet integrations, an extension of a previous method[5]. The fit parameters were osculating Jacobi orbital elements defined at dynamical epoch 2,455,190.0 (BJD): period $P$, the closest transit epoch $E_0$, the eccentricity vector components $e\cos\omega$ and $e\sin\omega$, and planetary mass $M_p$. These fits assumed zero mutual inclination and zero inclination to the line of sight, and they should be qualitatively valid for the small values expected (see SI Section 6). We found that Kepler-11g interacted very little with the other planets, consistent with its small transit timing deviations ($\sim 0.5$ min). At a nominal mass of a Neptune mass, it produces similarly small deviations in the other planets, so we abandoned hope of identifying the mass of Kepler-11g and fixed it, and its eccentricity, to zero in the fits.

Our first dynamical fit considered all the planets on circular orbits, so it had only 5 free parameters beyond the linear-ephemeris fits: the masses of the 5 inner planets. We found that all of the inner 5 planets have significantly detected masses. The resulting $\chi^2 = 110.34$ for 89 degrees of freedom (now 5 less for the five fitted planet masses; see Table S5 for the contributions from each planet), a formally acceptable fit to the data ($p$-value 6.2%). We use an F-test[45] to decide whether the new free parameters on the whole are justified. If the new free parameters are not statistically justified, the $\chi_i^2$ of the initial fit should follow a $\chi^2$ distribution of $\nu_i$ degrees of freedom ($\nu_i = 94$), and the $\chi_f^2$ of the final fit should follow a $\chi^2$ distribution of $\nu_f$ degrees of freedom ($\nu_f = 89$). Now we use the linearity property of $\chi^2$ distributions; the difference $\Delta\chi^2 \equiv \chi_i^2 - \chi_f^2$ should follow a $\chi^2$ distribution of $\Delta$d.o.f. $\equiv \nu_i - \nu_f$ degrees of freedom. We define the F-ratio is the ratio of (a) the improvement in $\chi^2$, normalized to the number of new free parameters:

$$\Delta\chi^2/\Delta\text{d.o.f.} = (191.51 - 110.34)/(94 - 89) = 16.23,$$

to (b) the final reduced $\chi^2$:

$$\chi_f^2/\nu_f = 110.34/89 = 1.23.$$

The F-ratio, a random variable composed of two random variables ($\Delta\chi^2$ and $\chi_f^2$), follows its own distribution. The F-test gives the probability ($p$-value) that the F-ratio is as high as it is by chance. In our case, the ratio of 13.20 has a $p$-value of $1 \times 10^{-9}$. Since the F-test compares the ratio of two reduced $\chi^2$ values, this calculation would have been the same if we had chosen to inflate the error bars at the beginning of the analysis, as both reduced $\chi^2$ values would decrease by an identical factor in response.

Since this dynamical fit was the simplest possible test of dynamical interactions being statistically present in the data, we regard it as a demonstration that dynamically significant transit time variations are actually present. The dynamical interactions are shown graphically in Figure S6a, and the fitted parameters are in Table S4. The F-test does not guarantee that any particular values of the 5 new free parameters are significant, but the associated formal errors suggest that each one is significant. The variation in each of the planets is linearly composed of perturbations by other planets (right hand panel of Figure S6a). The statistical reduction of each transiting planet's contribution to $\chi^2$ is given in Table S5.

One may wonder whether any signal of planetary orbital eccentricity may be extracted from the transit times. The first place to look is the inner two planets (Kepler-11b and c), as they have the most measured transit times and the closest relative spacing. We allowed their eccentricities to float, fitting $e\sin\omega$ and $e\cos\omega$, an additional 4 free parameters. These additional parameters resulted in a $\Delta\chi^2 = 13.41$ relative to the circular fit (Table S5). The $p$-value of an F-ratio of 13.41/4 compared to the final reduced $\chi^2$ of 96.93/85 is 2.5%, so the improvement is marginally significant. The fitted parameters are in Table S4.

We also found several distinct solutions in which planets all five of the closely-spaced planets Kepler-11b-f were allowed eccentric orbits; the Levenberg-Marquardt algorithm apparently did not find a unique, global minimum in this high-dimension space. The best $\chi^2$ was 85.48 for 79 degrees of freedom, which is not significantly better than the b/c eccentric Fit of Table S4 (F-test $p$-value of 12%). In the next subsection, we investigate stability. We found we needed to compromise between long-term stability and a good fit to the data. One solution that had all five inner planets eccentric — although only slightly eccentric, which is good for stability — is given in Table S4. Its $\chi^2 = 89.64$ (see Table S5), and the transit-timing model is shown in Figure S6b. The values of eccentricity for the outer planets, and the uniqueness of the solutions, will be subjects for follow-up work using (a) the short cadence data, (b) all observed transits including overlapping ones, and — most importantly — (c) more transits spanning a longer time baseline from *Kepler*.

All of these fits are statistically acceptable fits to the data, with quite Gaussian residuals. They, however, give slightly different best-fit masses. Therefore to derive masses in the main text, we compute the weighted mean mass from these three fits and adopt a generous error bar that spans the union of the $1\sigma$ intervals of all the fits.

### 4.1. Long-term stability of these solutions

We investigated long-term stability of the three solutions given in Table S4 using the hybrid integrator within the *Mercury* package[46], run on the supercomputer Pleiades at University of California, Santa Cruz. We set the switchover at 3 Hill radii, but in practice we aborted simulations that violated this limit, so for the bulk of the simulation the Mixed-Variable Symplectic method[47] was used, with a time step of 0.65124450 days. The simplest implementation[48] of general relativistic precession was used, an additional potential $U_{\rm GR} = -3(GM_\star/cr)^2$, where $G$ is Newton's constant, $c$ is the speed of light, and $r$ is the instantaneous distance from the star. More sophisticated methods[49] are not yet required, due to the uncertainties of the fits. These integrations used a stellar mass of 1.0 $M_\odot$. With respect to the best-fitting stellar mass of $0.95 \pm 0.10$ $M_\odot$, this choice implies slightly too fast precession due to relativity. We neglected precession due to tides or rotational flattening.

The simulations were run for a total of 250 Myr, a span for which the all-circular and all-eccentric fits survived. The b/c-eccentric solution became unstable at 169 Myr: after weakly chaotic jostling of eccentricities, a close encounter occurred between planets Kepler-11e and f (see Figure S7). However, we also ran an integration with very nearby initial conditions: one spatial coordinate displaced by only 1.5 meters. That system survived at least 250 Myr, despite showing similar weak chaos. Similar sensitivity of stability to initial conditions has been seen in other planetary systems with a high number of low-mass planets[50], and calculating stability maps of Kepler-11 would be useful future work.

The masses quoted in Table 1 of the main text can be used to calculate the number of mutual Hill sphere separations between pairs of planets. For the pairs (b–c, c–d, d–e, e–f), those separations are (7.0, 15.9, 10.9, 13.3), respectively. The criterion $\Delta \gtrsim 9$ applies only in situations in which all the separations between pairs equals the same $\Delta$. Therefore our integrations that show long-term stability are not in contradiction with previous work[20] on the stability of 3-or-more-planet systems.

The orbital eccentricity of Kepler-11b is expected to tidally damp on a timescale[30] of $\sim 0.5 - 5$ Gyr; the timescale for Kepler-11c is $\sim 0.2 - 20$ Gyr. The semi-major axis tidal decay rate could be non-zero for planets b and c, depending on tidal parameters. However, the ratio of semi-major axes and masses of these bodies are such that it is not clear (due to uncertainty in relative damping rates and errors in masses) whether these planets would have converging or diverging period ratios, which affects whether they will be caught into various mean-motion resonances as a result of tides in the planets.

## 5. Validation of Planet Kepler-11g

In the absence of dynamical confirmation (radial velocity variations or transit timing variations), validating the Kepler-11g signal as being of true planetary nature requires us to explore the enormous range of false positives that could mimic the signal. For this we use BLENDER, a technique[6] that models the *Kepler* light curve directly as a blend. Because of the extremely high precision of the *Kepler* photometry, BLENDER is able to place very tight constraints on the scenarios that can precisely reproduce the detailed shape of the transit signal. Further constraints are provided by our spectroscopic observations, by the photometry (color indices), and by the astrometry (centroid motion analysis). As described below, the combination of these allows us to rule out the vast majority of the possible contaminants for Kepler-11g. The remaining scenarios must be evaluated statistically.

In the BLENDER modeling, we refer to the target itself as the 'primary', and the contaminating pair of objects is composed of the 'secondary' and 'tertiary'. The intrinsic brightness of the primary needed for these simulations is based on the stellar parameters described previously, and the properties of the secondary and tertiary are derived from model isochrones[13,6]. We explore both hierarchical triple scenarios (an eclipsing star+star or star+planet pair physically associated with the target) and chance alignments (a spatially unresolved background or foreground pair of objects eclipsing each other).

We find that no hierarchical triple system with the tertiary being a star more massive than $0.1\,\mathrm{M}_\odot$ can mimic the observed light curve. When the tertiary is allowed to be a smaller object such as a planet, we do find one case of a hierarchical triple blend that cannot be ruled out, as it involves a secondary star that is faint enough that it would go undetected in our spectra. Based on the spectroscopic simulations described earlier in the Supplemental Information, and considering the signal-to-noise ratios of our spectra, for these simulations we adopt a simplified constraint such that any companions within 1 magnitude of the target are assumed to be detectable in our spectra. We consider this limit to be very conservative, as there are more sensitive ways of detecting spectroscopic companions than visual inspection of the cross-correlation functions, such as examination of the quality of the fit ($\chi^2$ statistic) of the spectral modeling with SME.

The blend scenarios involving star+planet pairs that are allowed by BLENDER contain a secondary star that can be significantly redder than Kepler-11, with a mass constrained to be between about 0.55 and 0.91 M$_\odot$ (see Figure S8). Secondaries below 0.55 M$_\odot$ (which are fainter than the target by 3.5 magnitudes or more) do not result in acceptable fits to the light curve, and companions over 0.91 M$_\odot$ would be bright enough ($\Delta Kp < 1$ mag) that we would usually see them in our spectra (green hatched region in the figure). Viable blends with secondaries in this range are orbited by a giant planetary companion (or a brown dwarf, or an extremely small star) in an eccentric orbit, with the transit occurring near apastron. In those cases, the slower orbital speed of the planet close to apastron (longer transit durations compared to a circular orbit; see Figure S9) allows for transits of a star smaller than Kepler-11 to reproduce the observed shape of the signal, within $3\sigma$ of the best Neptune-size transiting planet fit.

A large fraction of these hierarchical triple blends can be discriminated by comparing the predicted color of the blend with the observed color index of the *Kepler* target, which is $r - K_s = 1.473 \pm 0.036$ [ref. 37]. Any blends differing from this value by more than 0.11 mag are considered to be ruled out, at the $3\sigma$ level. This effectively excludes secondary masses between 0.58 and 0.85 M$_\odot$ (blue hatched region in Figure S8 and Figure S9), and further restricts the range of eccentricities and orientations allowed. Two narrow regions remain for secondaries between 0.55 and 0.58 M$_\odot$, and between 0.85 and 0.91 M$_\odot$. Blends with such secondaries orbited by a giant planet transiting near apastron could mimic the signal that we see, and would go undetected in our follow-up observations.

The frequency of this type of blend may be estimated by first calculating the fraction of stars in this portion of the *Kepler* field that have a binary companion with a secondary in the proper mass range, and then the fraction of those that are orbited by a giant planet (the fraction of stars with brown dwarf and very small stellar companions is much less) in a suitably eccentric orbit transiting near apastron. In order to compute this estimate, we simulated one million objects, assigning them a random orbital period and eccentricity drawn (with repetition) from the actual distributions seen in ground-based surveys[51]. With a random longitude of periastron drawn from a uniform distribution, we computed how many of them fall in the identified range of durations, and finally how many of them would be expected to transit. For a sampling of $10^6$ stars, we find that 11,094 have a binary companion in the proper mass range (assuming binary frequencies and a mass ratio distribution[52]). A total of 23.2 of those are expected to have a giant transiting planet (using a rate of occurrence from radial velocity studies[18], and accounting for the geometric transit probability), and in only 1.06 of these cases does the transit occur with eccentricities and orientations in the range allowed by BLENDER. Finally, 0.31 of these scenarios correspond to orbital periods between 47 days (the period of the neighboring planet interior to Kepler-11g) and 500 days (the approximate time span of the photometric observations). Thus, the expected frequency of blends due to hierarchical triples that are able to mimic the signal of Kepler-11g is $0.31 \times 10^{-6}$.

Next we considered blend scenarios involving chance alignments (background eclipsing binaries), with the eclipsing object being another star. As before, starting with $10^6$ simulated background stars, approximately 460,000 of them are expected to have a stellar companion[52]. Of these, 38,829 would have binary orbital periods between 47 and 500 days. On the other hand, only 4112 out of the 460,000 are expected to have orbits oriented such that the stars undergo eclipses. Combining the two constraints, we find that only 344 stars in the background of Kepler-11 would have an eclipsing companion in the period range considered. The constraints from BLENDER for this case are illustrated in Figure S10, where the relative distance between the binary and the *Kepler* target is parameterized in terms of the difference in distance modulus, for convenience. BLENDER places strong limits not only on the range of relative distances between the background binary and the target, but also on the mass of the secondary star (which in this case can be of solar type, or larger). Further constraints are provided by the centroid motion analysis described earlier, which rules out any stellar companions separated by more than 0.70 arc seconds from the target (Table S1). Additionally, the lack of double lines in our Keck spectra makes it unlikely that we have missed stars angularly closer than this within ∼1 magnitude of the target brightness. The combined constraints from BLENDER, centroid motion analysis, and spectroscopy imply that only a fraction 0.00078 of each background binary would be able to mimic the signal. Thus, we estimate the blend frequency for background eclipsing binaries to be $0.00078 \times 344 \times 10^{-6} = 0.27 \times 10^{-6}$.

Chance alignments with a star orbited by a transiting giant planet (as opposed to another smaller star) can also mimic the signal. Based on a recently determined occurrence rate[18], we expect that out of $10^6$ simulated stars, some 105,000 will be orbited by a giant planet, of which 30,059 will have periods between 47 and 500 days. On the other hand, only 2070 of the 105,000 are expected to transit their parent stars. When both effects are considered, we find that 153 out of the initial million stars will have a transiting giant planet in the proper period range. The BLENDER constraints for this case are illustrated in Figure S11. Both background and foreground blends are able to reproduce the transit light curve for this signal, and allowance for eccentric orbits leads to a wide range of spectral types (or masses) permitted for the secondary stars. But most of these scenarios are ruled out by other observations. In addition to the brightness limit from spectroscopy, the overall color of the blend is a strong discriminant, and is found to be inconsistent with the measured color of Kepler-11 in many of these cases, as shown in the figure. Folding together these constraints from BLENDER, spectroscopy, color information, as well as the angular separation limits mentioned earlier from the centroid motion analysis, we find that for every foreground/background star with a 47–500 day transiting giant planet, only 0.00205 are not excluded by any other type of observation. The blend frequency is therefore $0.00205 \times 153 \times 10^{-6} = 0.31 \times 10^{-6}$.

Adding together the blend frequencies for the three types of scenarios discussed above (hierarchical triples, and chance alignments with a star+star or star+planet pair), we find a total blend frequency of $(0.31 + 0.27 + 0.31) \times 10^{-6} = 0.89 \times 10^{-6}$.

Extrapolation is required to estimate the *a priori* probability of an outer planetary companion to Kepler-11, since observations do not directly constrain the frequency of companions as small and distant as the candidate that we are seeking to validate. We choose a nominal mass value of $10\,M_\oplus$, based upon the candidate's size and the size vs. mass relationship of the inner planets in the system (Table 1 of the main text). We estimate[18] that 1.9% of sunlike stars have a giant planet ($0.3\,M_J < M\sin i < 10\,M_J$) with period between 80 days (to exclude planets that might well be too close to Kepler-11f to be dynamically-stable) and 250 days (to give a high probability of there being at least two transits in our data set). Note that this period range is significantly narrower than the range we allowed for in our calculation of the *a priori* likelihood of false positives.

We are interested in the fraction with smaller planets, 10–100 $M_\oplus$, i.e., large enough to have transits likely detected, in this period range. That number has not been directly measured, but for periods shorter than 50 days, the mass dependence is $df/d\log M \propto M^{-0.48}$ over a range in masses that encompasses both giant planets and the intermediate planets of interest here[19], and we have no reason to think that this mass dependence is not approximately valid for the period range of relevance here. We find that 4.7% of stars should have a planet in the size-period range. For a random inclination distribution, the chance that a planet with period 118 days would transit Kepler-11 is 1.14%, yielding an overall *a priori* chance of a transiting planet in the mass and period range considered of $0.5 \times 10^{-3}$. If the inclination is not assumed to be random but is instead drawn from a Rayleigh distribution of mean 4° (based on the inner planets), the *a priori* probability of a transiting planet is an order of magnitude larger.

### 5.1. The Possibility of Multiple Planetary Systems

As in the case of Kepler-9 [ref. 5], mutual dynamical interactions show that planets b and c are in the same system and that planets Kepler-11d, e, and f are in the same system. The photometric analysis in combination with the BLENDER results certainly show no indication contrary to the hypothesis that all six of these planets are orbiting the same star. By considering qualitatively the alternative hypotheses, we reject them as simply too contrived compared to the much more likely case that all six planets simply orbit the same star.

Based on all the information available, the best alternative hypothesis is the blend of a wide binary in which each component is orbited by an edge-on planetary system. The component stars must have nearly the same spectrum, such that it is indistinguishable from a single star. For equal stars, all planetary transits would be diluted by a factor of two, indicating that the true planetary densities much less than the low densities we determined: $2^{-3/2} \approx 0.35$ times as large as the estimates given in Table 1 of the main text. For unequal stars, the planets orbiting the smaller star are diluted even more; and furthermore from the transit durations a denser secondary star would require near apocentric transits of multiple planets, which would raise additional probabilistic and stability concerns. Though extremely low planetary densities are not astrophysically impossible, they would considerably stretch our understanding of planet formation.

## 6. Mutual Inclination and Coplanarity

The observability of multiple transiting planets depends upon the individual on-sky inclinations of the component planets, which can be related in a probabilistic manner to the (much more physically interesting) mutual inclinations of their orbital planes with respect to one another[23]. In order to constrain the coplanarity of the planets in the Kepler-11 system, we performed several suites of Monte Carlo simulations over a range of differing mean mutual inclinations (MMI) and various configurations of the system: 1) the five inner planets only, 2) all six planets, and 3) all six observed planets plus a hypothetical planet with a period of 74.35 days, which is the geometric mean of the periods of Kepler-11f and g. We began each Monte Carlo realization by drawing a spherically isotropic on-sky inclination $i$, where $i = 90°$ is directly aligned with the observer's line of sight, therefore defining an arbitrary reference plane. We then populated the system with the planets of the observed Kepler-11 periods on circular orbits around a Sun-like star. For each planet we drew a mutual inclination with respect to the reference plane from a Rayleigh distribution with a Rayleigh parameter $\zeta$, such that the MMI $= \zeta(\pi/2)^{1/2}$. To determine if a given planet will transit with those orbital parameters it is necessary to transform the mutual inclination to an on-sky inclination. We accomplished this[23] by applying the spherical law of cosines:

$$\cos i_p = \cos i_{\rm ref} \cos\phi + \sin i_{\rm ref} \sin\phi \cos\delta \qquad (5)$$

where $i_p$, $i_{\rm ref}$ are the on-sky inclinations of the planet and reference plane, respectively, and $\phi$ is the mutual inclination between the planes. $\delta$ is a random angle that corresponds to the node of the planet's orbit on the reference plane. For each configuration of planets we also simulated the exactly coplanar case where $i_{\rm ref} = i_p$ for all the planets in the system.

Using the on-sky inclinations for each planet, we then calculated the separation between the center of the stellar disk and the center of the planet in the plane of the sky to determine whether the planet would transit for a given $i_p$ [ref. 53]. The planet was considered to transit if the value of this separation was less than one stellar radius when the planet was aligned on the observer's line of sight, i.e., the planet crossed the stellar disk. We then counted the number of planets seen to transit in each realization. To ensure high statistical precision we performed $5 \times 10^5$ realizations for each combination of planet configuration and MMI. The resulting transit probabilities of these simulations are presented in Table S6.

For circular orbits that are exactly coplanar, the probability of $N$ planets transiting is just the geometric probability of the $N^{\rm th}$ planet transiting given by the ratio of the star's radius and the semimajor axis of the $N^{\rm th}$ planet, i.e., if the outermost planet transits, then all the interior planets must also transit. This gives an upper limit of 2.0% to the transit probability for the five inner planets and 1.1% for all six planets, which is in very good agreement with the results from the coplanar simulations and the trials with the MMI set to $0.001°$. If the actual transit duration ($D(b)$) of a particular planet had to be at least $X_{\rm min}$ times as long as a central transit ($D(b=0)$) in order for it to have been detected from existing data, then the probability of detecting that planet decreases. It is as if the effective star of the size is reduced from $R_\star$ to $b_{\rm max} \times R_\star$. For

– 14 –

$X_{\min} = 0.5$, $b_{\max} = 0.86$. If a small planet resided between Kepler-11f and g, then the probability that its transits and would have been detected from the present data set is less than indicated in Table S6. Of course, the minimum detectable duration depends on the planet size, complicating the interpretation of Monte Carlo simulations.

All of the planet configurations explored demonstrate that, within the statistical error, the coplanar case maximizes the probability for multiple transits, as one would expect, except when there is a hypothetical planet between Kepler-11f and g. In these configurations, the seventh planet significantly decreases the probability of observing only six planets when the orbits are very nearly coplanar. Conversely, to duplicate the observed planets of Kepler-11, it requires at least some spread in $i_p$ in order for the seventh planet to transit without also observing the sixth. The probability for both these cases is highest with a MMI of $\sim 0.8°$. When considering only the five inner planets or all six planets without additional unseen planets, they have relatively high transit probabilities of $\sim 1.0\%$ for similar MMIs. In all cases, however, the probability of observing six transiting planets becomes small by a MMI of $3.0°$, and for observing the inner five planets the probability drops off by $4.0°$. Thus the results of these simulations strongly suggest a MMI for the Kepler-11 system of $\sim 1.0°$, which is somewhat smaller than the mean inclination for the planets of the Solar System of $2.32°$ with respect to the ecliptic plane.

Mutual inclinations could also be determined from exoplanet mutual events, where one planet crosses over another in the plane of the sky[23]. During the course of these observations, several doubly-concurrent transits are seen with one triply-concurrent transit (Figure S12). In none of these cases is there evidence for a mutual event (i.e., an overlapping double transit); in any case, the mutual event signal would be quite small.

**Acknowledgments:** Chris Henze assisted in performing numerical simulations with BLENDER at the NASA Advanced Supercomputing Division (NASA Ames Research Center). This work is based in part on data obtained as part of the UKIRT Infrared Deep Sky Survey.

## REFERENCES

[35] Bryson, S. T. *et al.* The Kepler Pixel Response Function. *Astrophys. J.* **713**, L97–L102 (2010).

[36] Bryson, S. T. *et al.* Selecting Pixels for Kepler Downlink. *Proc. SPIE* **7740**, 77401D (2010).

[37] Latham, D. W. *et al.* The Kepler Input Catalog. *Bull. Am. Astron. Soc.* **37**, 1340 (2005)

[38] Orosz, J. A. & Hauschildt, P. H. The use of the NextGen model atmospheres for cool giants in a light curve synthesis code. *Astron. Astrophys.* **364**, 265–281 (2000).

[39] Charbonneau, P. Genetic Algorithms in Astronomy and Astrophysics. *Astrophys. J. Suppl.* **101**, 309–334 (1995).

[40] Tegmark, M. *et al.* Cosmological parameters from SDSS and WMAP. *Phys. Rev. D* **69**, 103501 (2004).

[41] Giménez, A. Equations for the analysis of the light curves of extra-solar planetary transits. *Astron. Astrophys.* **450**, 1231–1237 (2006).

[42] http://astro4.ast.villanova.edu/aprsa/?q=node/8

[43] Zechmeister, M. & Kürster, M. The generalised Lomb-Scargle periodogram. A new formalism for the floating-mean and Keplerian periodograms. *Astron. Astrophys.* **496**, 577–584 (2009).

[44] Nesvorný, D. & Morbidelli, A. Mass and Orbit Determination from Transit Timing Variations of Exoplanets. *Astrophys. J.* **688**, 636–646 (2008).

[45] Bevington, P. R. & Robinson, D. K. *Data reduction and error analysis for the physical sciences.* 2nd Edition, (McGraw-Hill 1992).

[46] Chambers. J. E. A hybrid symplectic integrator that permits close encounters between massive bodies. *Mon. Not. R. Astron. Soc.* **304**, 793–799 (1999).

[47] Wisdom, J. & Holman, M. Symplectic maps for the n-body problem. *Astron. J.* **102**, 1528–1538 (1991).

[48] Nobili, A. & Roxburgh, I. W. Simulation of general relativistic corrections in long term numerical integrations of planetary orbits. In *Celestial Mechanics and Astrometry. High Precision Dynamical Theories and Observational Verifications* (eds Kovalevsky, J. & Brumberg, V. A.), IAU Symp. **114**, 105–110 (1986)

[49] Saha, P. & Tremaine, S. Long-term planetary integration with individual time steps. *Astron. J.* **108**, 1962–1969 (1994).

[50] Lovis, C. *et al.* The HARPS search for southern extra-solar planets. XXVII. Up to seven planets orbiting HD 10180: probing the architecture of low-mass planetary systems. Preprint at ⟨http://arXiv.org/abs/1011.4994⟩ (2010).

[51] http://exoplanets.eu

[52] Raghavan, D. *et al.* A Survey of Stellar Families: Multiplicity of Solar-type Stars. *Astrophys. J. Suppl. Ser.* **190**, 1–42 (2010).

[53] Kipping. D. M. Investigations of approximate expressions for the transit duration. *Mon. Not. R. Astron. Soc.* **407**, 301–313 (2010).

Table S1. Results of the centroid analysis.

| Planet | Offset (arc seconds) |
|---|---|
| b | 0.41 ± 0.46 |
| c | 0.27 ± 0.21 |
| d | 0.08 ± 0.22 |
| e | 0.36 ± 0.18 |
| f | 0.40 ± 0.43 |
| g | 0.10 ± 0.23 |

Table S2a. Linear ephemerides for planets in the Kepler-11 system.

| Planet | Period (days) | Epoch (BJD) |
|---|---|---|
| b | 10.30375 ± 0.00016 | 2,454,971.5052 ± 0.0077 |
| c | 13.02502 ± 0.00008 | 2,454,971.1748 ± 0.0031 |
| d | 22.68719 ± 0.00021 | 2,454,981.4550 ± 0.0044 |
| e | 31.99590 ± 0.00028 | 2,454,987.1590 ± 0.0037 |
| f | 46.68876 ± 0.00074 | 2,454,964.6487 ± 0.0059 |
| g | 118.37774 ± 0.00112 | 2,455,120.2901 ± 0.0022 |

Note. — This ephemeris establishes the calculated ("C") times from which transit timing measurements are referenced below. Uncertainty in epoch is median absolute deviation of transit times from this ephemeris; uncertainty in period is this quantity divided by number of orbits between first and last observed transits.

Table S2b.   Transit times for Kepler-11b.

| $N$ | Observed<br>(BJD − 2,455,000) | $O - C$<br>(days) | Uncertainty<br>(days) |
|---|---|---|---|
| 0  | −28.4971 | −0.0023 | 0.0070 |
| 1  | −18.1754 | +0.0156 | 0.0088 |
| 2  | −7.8849  | +0.0024 | 0.0079 |
| 4  | 12.7234  | +0.0032 | 0.0109 |
| 5  | 23.0214  | −0.0026 | 0.0099 |
| 6  | 33.3342  | +0.0065 | 0.0081 |
| 7  | 43.6414  | +0.0100 | 0.0098 |
| 10 | 74.5570  | +0.0143 | 0.0115 |
| 11 | 84.8410  | −0.0054 | 0.0102 |
| 12 | 95.1267  | −0.0235 | 0.0089 |
| 13 | 105.4402 | −0.0137 | 0.0084 |
| 14 | 115.7453 | −0.0123 | 0.0075 |
| 15 | 126.0537 | −0.0077 | 0.0104 |
| 16 | 136.3701 | +0.0050 | 0.0115 |
| 17 | 146.6638 | −0.0051 | 0.0109 |
| 18 | 156.9669 | −0.0058 | 0.0121 |
| 19 | 167.2670 | −0.0094 | 0.0092 |
| 21 | 187.8976 | +0.0137 | 0.0112 |
| 25 | 229.1093 | +0.0105 | 0.0105 |
| 28 | 260.0409 | +0.0308 | 0.0083 |
| 29 | 270.3241 | +0.0102 | 0.0106 |
| 30 | 280.6209 | +0.0033 | 0.0079 |
| 31 | 290.9219 | +0.0006 | 0.0104 |
| 32 | 301.2332 | +0.0081 | 0.0071 |
| 33 | 311.5136 | −0.0152 | 0.0096 |
| 35 | 332.1403 | +0.0040 | 0.0101 |
| 36 | 342.4343 | −0.0058 | 0.0071 |
| 37 | 352.7263 | −0.0175 | 0.0077 |
| 38 | 363.0409 | −0.0067 | 0.0109 |
| 39 | 373.3537 | +0.0024 | 0.0168 |
| 40 | 383.6558 | +0.0007 | 0.0170 |
| 42 | 404.2598 | −0.0027 | 0.0157 |
| 43 | 414.5443 | −0.0220 | 0.0153 |
| 44 | 424.8755 | +0.0055 | 0.0157 |
| 46 | 445.4921 | +0.0146 | 0.0168 |
| 47 | 455.7985 | +0.0173 | 0.0112 |

Table S2c. Transit times for Kepler-11c.

| $N$ | Observed (BJD − 2,455,000) | $O - C$ (days) | Uncertainty (days) |
|---|---|---|---|
| 0 | −28.8280 | −0.0028 | 0.0052 |
| 1 | −15.7977 | +0.0025 | 0.0036 |
| 2 | −2.7768 | −0.0017 | 0.0047 |
| 3 | 10.2396 | −0.0103 | 0.0040 |
| 4 | 23.2713 | −0.0035 | 0.0038 |
| 5 | 36.2991 | −0.0008 | 0.0050 |
| 7 | 62.3541 | +0.0041 | 0.0036 |
| 8 | 75.3781 | +0.0031 | 0.0036 |
| 10 | 101.4359 | +0.0109 | 0.0042 |
| 12 | 127.4731 | −0.0019 | 0.0044 |
| 14 | 153.5287 | +0.0037 | 0.0041 |
| 15 | 166.5509 | +0.0008 | 0.0040 |
| 16 | 179.5659 | −0.0092 | 0.0042 |
| 17 | 192.5998 | −0.0003 | 0.0042 |
| 18 | 205.6148 | −0.0103 | 0.0073 |
| 22 | 257.7202 | −0.0051 | 0.0038 |
| 23 | 270.7516 | +0.0014 | 0.0040 |
| 24 | 283.7789 | +0.0036 | 0.0034 |
| 25 | 296.8069 | +0.0066 | 0.0033 |
| 26 | 309.8308 | +0.0055 | 0.0043 |
| 27 | 322.8503 | −0.0001 | 0.0039 |
| 28 | 335.8780 | +0.0026 | 0.0031 |
| 29 | 348.9022 | +0.0018 | 0.0034 |
| 30 | 361.9203 | −0.0051 | 0.0039 |
| 31 | 374.9493 | −0.0011 | 0.0058 |
| 32 | 387.9698 | −0.0057 | 0.0057 |
| 33 | 400.9995 | −0.0009 | 0.0069 |
| 34 | 414.0261 | +0.0006 | 0.0066 |
| 35 | 427.0378 | −0.0127 | 0.0062 |
| 36 | 440.0669 | −0.0086 | 0.0090 |
| 37 | 453.0987 | −0.0019 | 0.0069 |

Table S2d.  Transit times for Kepler-11d.

| N | Observed (BJD − 2,455,000) | $O - C$ (days) | Uncertainty (days) |
|---|---|---|---|
| 0 | −18.5423 | +0.0027 | 0.0034 |
| 1 | 4.1417 | −0.0004 | 0.0030 |
| 2 | 26.8291 | −0.0002 | 0.0034 |
| 4 | 72.2067 | +0.0030 | 0.0037 |
| 6 | 117.5731 | −0.0049 | 0.0038 |
| 8 | 162.9474 | −0.0050 | 0.0039 |
| 9 | 185.6353 | −0.0044 | 0.0035 |
| 12 | 253.7130 | +0.0118 | 0.0029 |
| 14 | 299.0691 | −0.0065 | 0.0038 |
| 15 | 321.7581 | −0.0046 | 0.0047 |
| 16 | 344.4405 | −0.0095 | 0.0032 |
| 17 | 367.1425 | +0.0054 | 0.0036 |
| 18 | 389.8269 | +0.0026 | 0.0050 |
| 19 | 412.5134 | +0.0019 | 0.0054 |
| 21 | 457.8903 | +0.0044 | 0.0051 |

Table S2e.  Transit times for Kepler-11e.

| N | Observed (BJD − 2,455,000) | $O - C$ (days) | Uncertainty (days) |
|---|---|---|---|
| 0 | −12.8460 | −0.0049 | 0.0021 |
| 1 | 19.1520 | −0.0029 | 0.0020 |
| 2 | 51.1543 | +0.0035 | 0.0020 |
| 3 | 83.1484 | +0.0017 | 0.0022 |
| 4 | 115.1465 | +0.0039 | 0.0019 |
| 5 | 147.1421 | +0.0037 | 0.0023 |
| 6 | 179.1359 | +0.0016 | 0.0029 |
| 7 | 211.1275 | −0.0027 | 0.0025 |
| 8 | 243.1200 | −0.0061 | 0.0024 |
| 10 | 307.1167 | −0.0012 | 0.0024 |
| 11 | 339.1200 | +0.0062 | 0.0024 |
| 13 | 403.0945 | −0.0111 | 0.0040 |

Table S2f. Transit times for Kepler-11f.

| $N$ | Observed<br>(BJD − 2,455,000) | $O - C$<br>(days) | Uncertainty<br>(days) |
|---|---|---|---|
| 1 | 11.3450 | +0.0075 | 0.0071 |
| 2 | 58.0266 | +0.0004 | 0.0050 |
| 3 | 104.7189 | +0.0040 | 0.0056 |
| 4 | 151.3967 | −0.0070 | 0.0069 |
| 5 | 198.0769 | −0.0156 | 0.0076 |
| 6 | 244.7754 | −0.0059 | 0.0112 |
| 7 | 291.4669 | −0.0031 | 0.0072 |
| 8 | 338.1590 | +0.0003 | 0.0078 |
| 9 | 384.8713 | +0.0238 | 0.0102 |

Table S2g. Transit times for Kepler-11g.

| $N$ | Observed<br>(BJD − 2,455,000) | $O - C$<br>(days) | Uncertainty<br>(days) |
|---|---|---|---|
| 0 | 120.2884 | −0.0017 | 0.0031 |
| 1 | 238.6714 | +0.0036 | 0.0032 |
| 2 | 357.0433 | −0.0022 | 0.0036 |

Table S3. Transit Model Fits.

| Planet | b | c | d | e | f | g |
|---|---|---|---|---|---|---|
| Transit depth (percent) | 0.031 ± 0.001 | 0.082 ± 0.001 | 0.098 ± 0.002 | 0.140 ± 0.002 | 0.055 ± 0.002 | 0.115 ± 0.003 |
| Transit duration (hr) | 4.02 ± 0.08 | 4.62 ± 0.05 | 5.58 ± 0.06 | 4.33 ± 0.08 | 6.54 ± 0.14 | 9.60 ± 0.16 |
| $R_p/R_\star$ | 0.01638 ± 0.00054 | 0.02615 ± 0.00064 | 0.02861 ± 0.00070 | 0.03791 ± 0.00095 | 0.02171 ± 0.00069 | 0.03087 ± 0.00080 |
| Impact parameter $b$ | $0.46^{+0.14}_{-0.28}$ | $0.36^{+0.16}_{-0.36}$ | $0.35^{+0.18}_{-0.35}$ | $0.79^{+0.04}_{-0.05}$ | $0.49^{+0.12}_{-0.21}$ | $0.35^{+0.17}_{-0.34}$ |
| $R_p$ ($R_\oplus$) | 1.97 ± 0.19 | 3.15 ± 0.30 | 3.43 ± 0.32 | 4.52 ± 0.43 | 2.61 ± 0.25 | 3.66 ± 0.35 |
| median uncertainty (mag) | 0.0002072 | 0.0002014 | 0.0002028 | 0.0002117 | 0.0002000 | 0.0001956 |
| No contamination assumed | | | | | | |
| Transit depth (percent) | 0.030 ± 0.001 | 0.078 ± 0.001 | 0.093 ± 0.002 | 0.133 ± 0.002 | 0.052 ± 0.002 | 0.108 ± 0.002 |
| Transit duration (hr) | 4.01 ± 0.09 | 4.61 ± 0.05 | 5.57 ± 0.06 | 4.32 ± 0.08 | 6.53 ± 0.15 | 9.60 ± 0.13 |
| $R_p/R_\star$ | 0.01595 ± 0.00052 | 0.02556 ± 0.00063 | 0.02787 ± 0.00065 | 0.03701 ± 0.00088 | 0.02121 ± 0.00069 | 0.03011 ± 0.00077 |
| Impact parameter $b$ | $0.46^{+0.14}_{-0.29}$ | $0.39^{+0.13}_{-0.39}$ | $0.35^{+0.18}_{-0.35}$ | $0.79^{+0.04}_{-0.06}$ | $0.49^{+0.12}_{-0.22}$ | $0.40^{+0.14}_{-0.40}$ |
| $R_p$ ($R_\oplus$) | 1.91 ± 0.18 | 3.12 ± 0.29 | 3.36 ± 0.32 | 4.42 ± 0.42 | 2.54 ± 0.25 | 3.65 ± 0.34 |
| median uncertainty (mag) | 0.0001956 | 0.0001918 | 0.0001931 | 0.0002021 | 0.0001909 | 0.0001840 |
| 10% contamination assumed | | | | | | |
| Transit depth (percent) | 0.033 ± 0.001 | 0.086 ± 0.001 | 0.104 ± 0.002 | 0.147 ± 0.003 | 0.058 ± 0.002 | 0.120 ± 0.003 |
| Transit duration (hr) | 4.02 ± 0.09 | 4.62 ± 0.05 | 5.58 ± 0.06 | 4.34 ± 0.08 | 6.54 ± 0.15 | 9.65 ± 0.14 |
| $R_p/R_\star$ | 0.01682 ± 0.00058 | 0.02681 ± 0.00067 | 0.02937 ± 0.00070 | 0.03893 ± 0.00098 | 0.02235 ± 0.00074 | 0.03174 ± 0.00087 |
| Impact parameter $b$ | $0.46^{+0.13}_{-0.29}$ | $0.34^{+0.20}_{-0.34}$ | $0.33^{+0.20}_{-0.33}$ | $0.79^{+0.04}_{-0.06}$ | $0.49^{+0.12}_{-0.23}$ | $0.39^{+0.14}_{-0.39}$ |
| $R_p$ ($R_\oplus$) | 2.02 ± 0.20 | 3.21 ± 0.30 | 3.50 ± 0.33 | 4.62 ± 0.44 | 2.67 ± 0.26 | 3.86 ± 0.37 |
| median uncertainty (mag) | 0.0002195 | 0.0002131 | 0.0002146 | 0.0002245 | 0.0002122 | 0.0002044 |

Table S4. Dynamical Fits.

| Planet | $P$ (days) | $T_0$ | $e\cos\omega$ | $e\sin\omega$ | $M_p/M_\star$ ($\times 10^{-6}$) |
|---|---|---|---|---|---|
| | | | Circular Fit | | |
| b | **10.3045** | **187.8971** | 0 | 0 | **16** |
| | ±0.0003 | ±0.0024 | ⋯ | ⋯ | ±3 |
| c | **13.0247** | **192.5953** | 0 | 0 | **50** |
| | ±0.0002 | ±0.0013 | ⋯ | ⋯ | ±7 |
| d | **22.6849** | **185.6366** | 0 | 0 | **18** |
| | ±0.0007 | ±0.0011 | ⋯ | ⋯ | ±4 |
| e | **32.0001** | **179.1365** | 0 | 0 | **26** |
| | ±0.0008 | ±0.0009 | ⋯ | ⋯ | ±5 |
| f | **46.6908** | **198.0844** | 0 | 0 | **6** |
| | ±0.0010 | ±0.0030 | ⋯ | ⋯ | ±3 |
| g | **118.3808** | **238.6688** | 0 | 0 | 0 |
| | ±0.0025 | ±0.0019 | ⋯ | ⋯ | ⋯ |
| | | | b/c Eccentric Fit | | |
| Planet | $P$ (days) | $T_0$ | $e\cos\omega$ | $e\sin\omega$ | $M_p/M_\star$ ($\times 10^{-6}$) |
| b | **10.3063** | **187.8927** | **0.0534** | **−0.0039** | **11** |
| | ±0.0007 | ±0.0028 | ±0.0383 | ±0.0072 | ±4 |
| c | **13.0241** | **192.5971** | **0.0416** | **−0.0007** | **34** |
| | ±0.0004 | ±0.0014 | ±0.0332 | ±0.0060 | ±11 |
| d | **22.6829** | **185.6365** | 0 | 0 | **18** |
| | ±0.0017 | ±0.0012 | ⋯ | ⋯ | ±4 |
| e | **32.0001** | **179.1363** | 0 | 0 | **26** |
| | ±0.0009 | ±0.0010 | ⋯ | ⋯ | ±5 |
| f | **46.6909** | **198.0844** | 0 | 0 | **7** |
| | ±0.0010 | ±0.0030 | ⋯ | ⋯ | ±3 |
| g | **118.3805** | **238.6687** | 0 | 0 | 0 |
| | ±0.0025 | ±0.0019 | ⋯ | ⋯ | ⋯ |
| | | | All-Eccentric Fit | | |
| Planet | $P$ (days) | $T_0$ | $e\cos\omega$ | $e\sin\omega$ (*) | $M_p/M_\star$ ($\times 10^{-6}$) |
| b | **10.3062** | **187.8939** | **0.0030** | **−0.0009** | **12** |
| | ±0.0007 | ±0.0039 | ±0.0088 | ±0.0026 | ±5 |
| c | **13.0240** | **192.5968** | **−0.0026** | **0.0011** | **36** |
| | ±0.0004 | ±0.0017 | ±0.0078 | ±0.0022 | ±11 |
| d | **22.6823** | **185.6367** | **−0.0127** | **0.0148** | **23** |
| | ±0.0014 | ±0.0020 | ±0.0261 | ±0.0064 | ±6 |
| e | **32.0027** | **179.1368** | **−0.0161** | **0.000005** | **28** |
| | ±0.0021 | ±0.0014 | ±0.0200 | ±0.000020 | ±7 |
| f | **46.6908** | **198.0837** | **−0.0119** | **−0.0037** | **10** |
| | ±0.0033 | ±0.0038 | ±0.0203 | ±0.0090 | ±5 |
| g | **118.3812** | **238.6690** | 0 | 0 | 0 |
| | ±0.0029 | ±0.0023 | ⋯ | ⋯ | ⋯ |

Note. — Osculating Jacobian orbital elements at dynamical epoch 2,455,190 (BJD). Transit epoch times $T_0$ are BJD − 2,455,000.
(*) The formal errors the Levenberg-Marquardt algorithm returned are exceedingly small for this parameter, for unknown reasons.

Table S5. $\chi^2$ per degrees of freedom.

| Planet | Constant-period | All-circular fit | b/c Eccentric fit | All-eccentric fit |
|--------|-----------------|------------------|-------------------|-------------------|
| b      | 54.50 / 34      | 38.92 / 33       | 27.67 / 31        | 28.32 / 31        |
| c      | 42.52 / 29      | 29.72 / 28       | 29.58 / 26        | 29.63 / 26        |
| d      | 38.94 / 13      | 20.16 / 12       | 18.12 / 12        | 16.67 / 10        |
| e      | 40.80 / 10      | 14.09 / 9        | 14.08 / 9         | 8.19 / 7          |
| f      | 12.83 / 7       | 5.64 / 6         | 5.66 / 6          | 5.10 / 4          |
| g      | 1.93 / 1        | 1.80 / 1         | 1.81 / 1          | 1.73 / 1          |
| total  | 191.51 / 94     | 110.34 / 89      | 96.93 / 85        | 89.64 / 79        |

Note. — The linear ephemeris (constant-period) model is compared here with the dynamical fits to derive planetary masses: the all-circular fit, the b/c eccentric fit, and the all-eccentric fit of Table S4. In each cell we list the contribution of each planet to the total $\chi^2$ and to the number of degrees of freedom [number of transit times for that planet, minus 2 (since period and epoch is always fit), and minus 1 (when mass alone is fit) or minus 3 (when mass and two eccentricity components are fit)].

Table S6.  Probabilities of multiple planet transits as a function of the mean mutual inclination, from Monte Carlo simulations.

| MMI (°) | 5 Planets (%) | 6 Planets (%) | 6 Planets[a] (%) | Kepler-11[a] (%) |
|---|---|---|---|---|
| 0.0    | $2.047 \pm 0.039$ | $1.109 \pm 0.029$ | $0.419 \pm 0.018$ | 0.000 |
| 0.0001 | $2.044 \pm 0.039$ | $1.098 \pm 0.029$ | $0.397 \pm 0.017$ | 0.000 |
| 0.05   | $2.040 \pm 0.039$ | $1.085 \pm 0.029$ | $0.402 \pm 0.018$ | 0.000 |
| 0.1    | $2.055 \pm 0.039$ | $1.095 \pm 0.029$ | $0.403 \pm 0.018$ | $0.002 \pm 0.001$ |
| 0.2    | $2.022 \pm 0.039$ | $1.098 \pm 0.029$ | $0.438 \pm 0.018$ | $0.029 \pm 0.005$ |
| 0.4    | $1.933 \pm 0.038$ | $1.073 \pm 0.029$ | $0.606 \pm 0.022$ | $0.131 \pm 0.010$ |
| 0.6    | $1.800 \pm 0.037$ | $0.911 \pm 0.026$ | $0.734 \pm 0.024$ | $0.194 \pm 0.012$ |
| 0.8    | $1.535 \pm 0.034$ | $0.767 \pm 0.024$ | $0.827 \pm 0.025$ | $0.232 \pm 0.013$ |
| 1.0    | $1.368 \pm 0.032$ | $0.591 \pm 0.021$ | $0.811 \pm 0.025$ | $0.225 \pm 0.013$ |
| 1.5    | $0.843 \pm 0.025$ | $0.285 \pm 0.015$ | $0.574 \pm 0.021$ | $0.151 \pm 0.011$ |
| 2.0    | $0.502 \pm 0.020$ | $0.128 \pm 0.010$ | $0.318 \pm 0.016$ | $0.085 \pm 0.008$ |
| 2.5    | $0.279 \pm 0.015$ | $0.066 \pm 0.007$ | $0.181 \pm 0.012$ | $0.040 \pm 0.006$ |
| 3.0    | $0.169 \pm 0.011$ | $0.029 \pm 0.005$ | $0.091 \pm 0.008$ | $0.024 \pm 0.004$ |
| 3.5    | $0.103 \pm 0.009$ | $0.018 \pm 0.004$ | $0.057 \pm 0.007$ | $0.012 \pm 0.003$ |
| 4.0    | $0.071 \pm 0.007$ | $0.011 \pm 0.003$ | $0.031 \pm 0.005$ | $0.006 \pm 0.002$ |
| 4.5    | $0.047 \pm 0.006$ | $0.004 \pm 0.002$ | $0.019 \pm 0.004$ | $0.005 \pm 0.002$ |
| 5.0    | $0.028 \pm 0.005$ | $0.004 \pm 0.002$ | $0.015 \pm 0.003$ | $0.004 \pm 0.002$ |
| 5.5    | $0.022 \pm 0.004$ | $0.002 \pm 0.001$ | $0.009 \pm 0.003$ | $0.002 \pm 0.001$ |
| 6.0    | $0.014 \pm 0.003$ | 0.001             | $0.004 \pm 0.002$ | 0.001 |
| 8.0    | $0.007 \pm 0.002$ | 0.000             | $0.002 \pm 0.001$ | 0.000 |
| 10.0   | $0.003 \pm 0.001$ | 0.000             | 0.000             | 0.000 |

[a]Includes a hypothetical 7$^{\text{th}}$ planet at $P = 74.35$ days.

Note. — Column 1 (MMI) is the mean mutual inclination; column 2 is the probability of the five inner planets, Kepler-11b–f, being observed to transit the host star; column 3 is the probability that all six planets transit; columns 4 and 5 are the result of adding a hypothetical planet with $P = 74.35$ days to the system. Column 4 corresponds to the probability of observing six transits in any combination, while column 5 is the probability that Kepler-11 obtains, i.e., the hypothetical planet is unobserved but we see planets b through g transit. The quoted error is the 95% confidence interval estimated by normal approximation where appropriate.

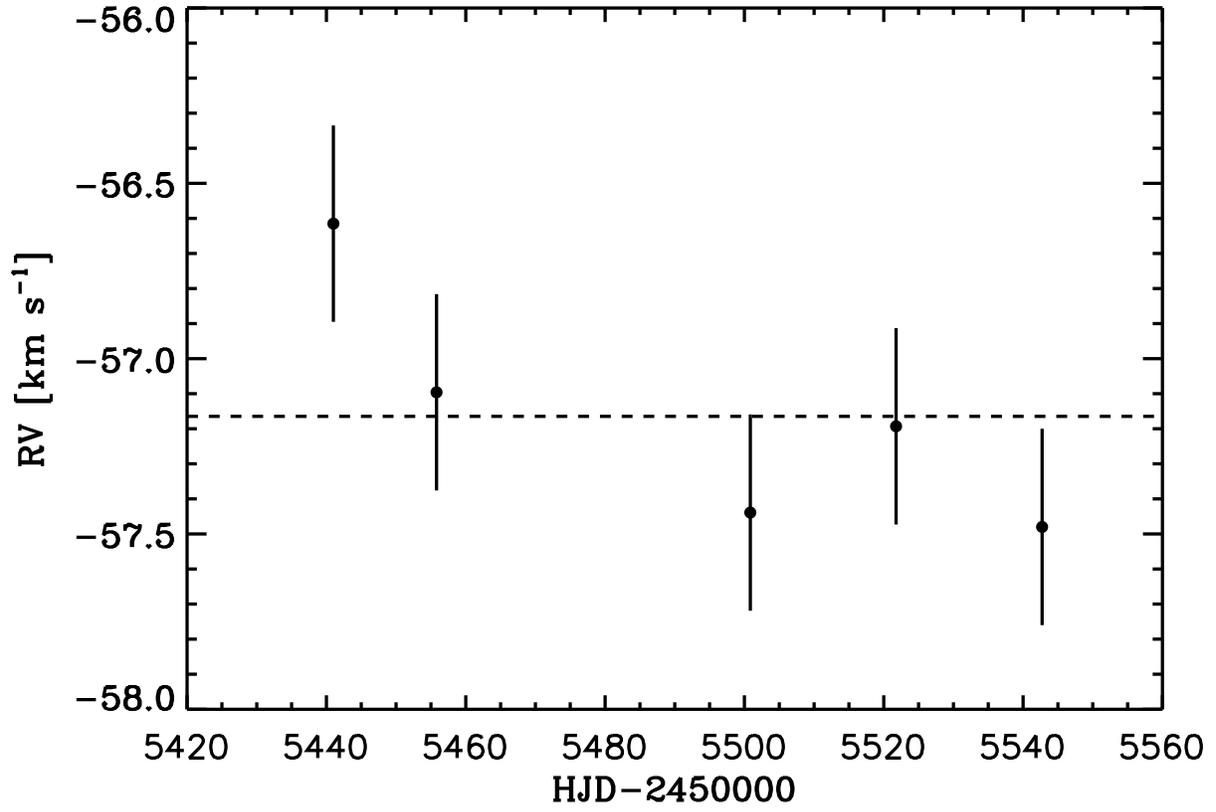

Fig. S1.— Radial velocities of the star Kepler-11.

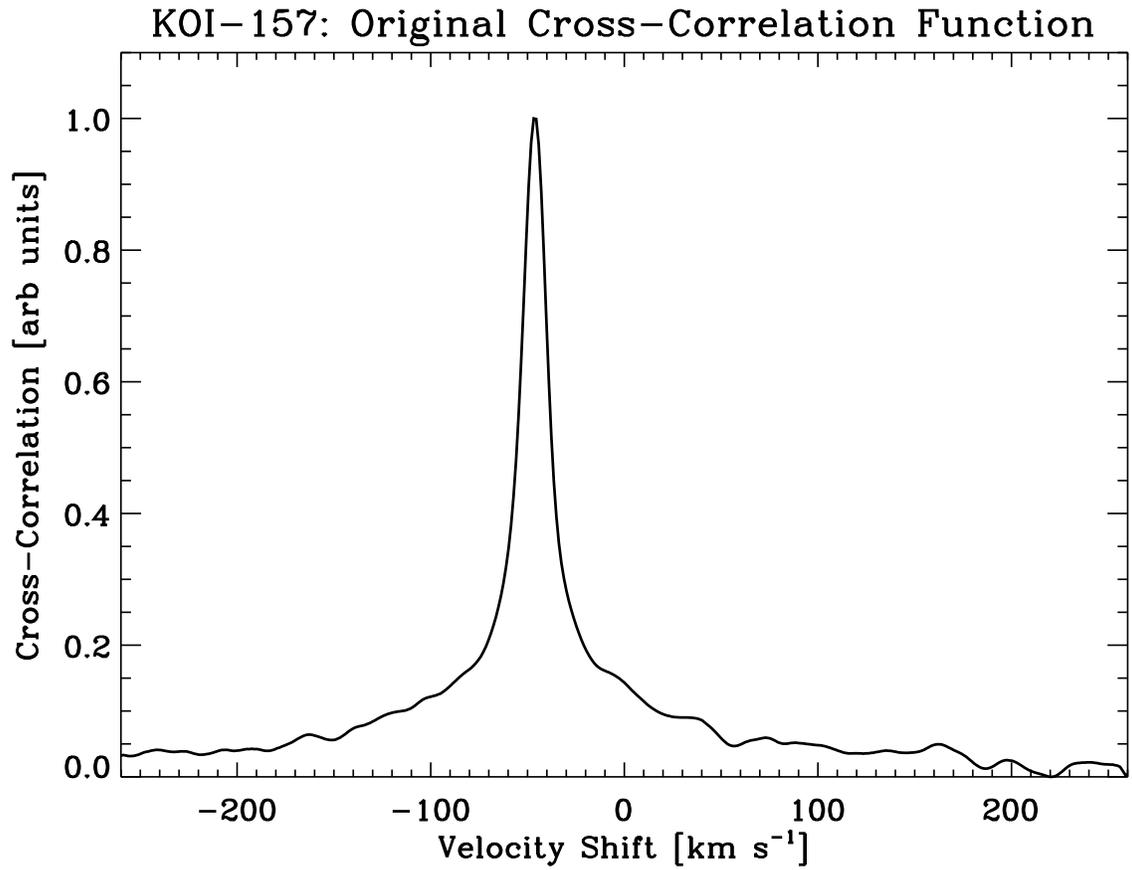

Fig. S2.— Cross-correlation function of the spectrum of Kepler-11 vs. the Sun's spectrum. The sharp, symmetric cross-correlation function suggests a single Sun-like star.

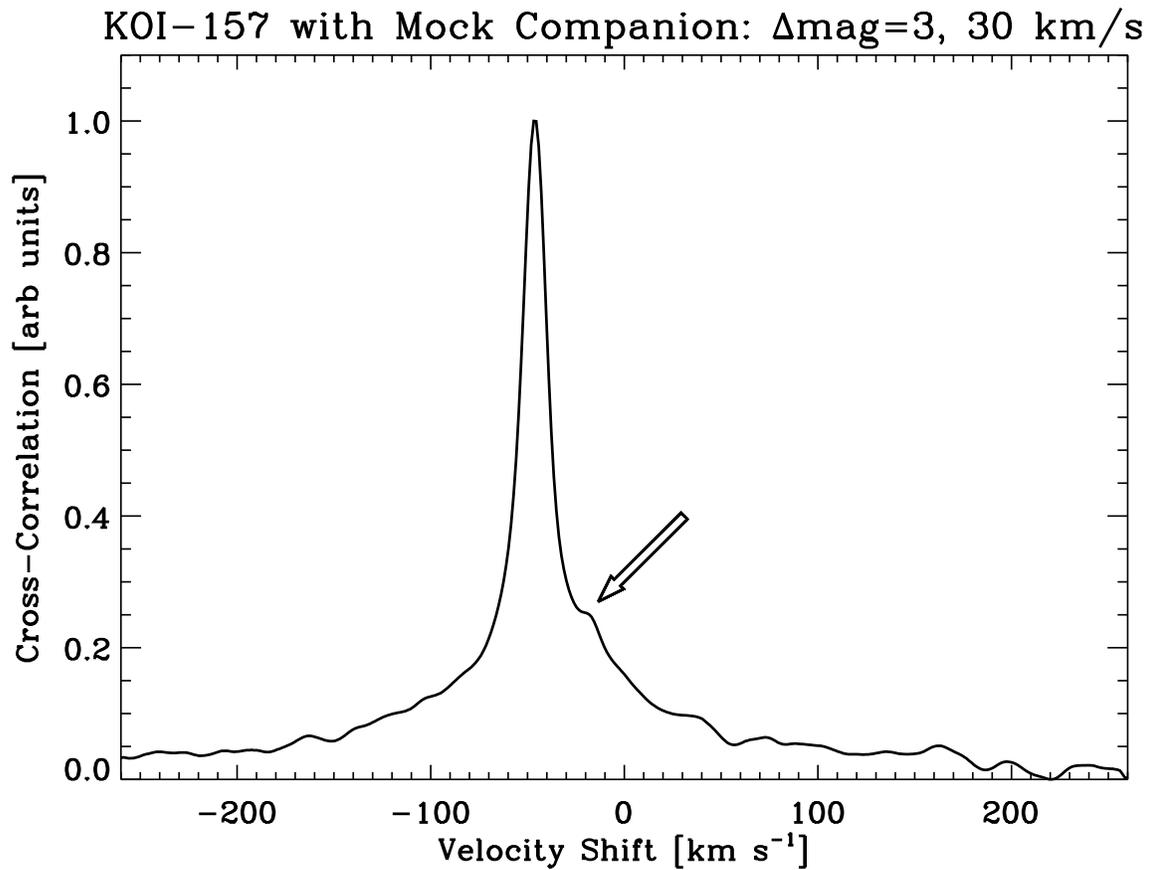

Fig. S3.— Simulated cross-correlation function for the spectrum of Kepler-11. The spectrum of another G star three magnitudes fainter than the target star with a radial velocity differing by $30\,\mathrm{km\,s^{-1}}$ is added. The original cross-correlation function of the Keck spectrum of Kepler-11 shows no evidence of a "bump" from a stellar companion.

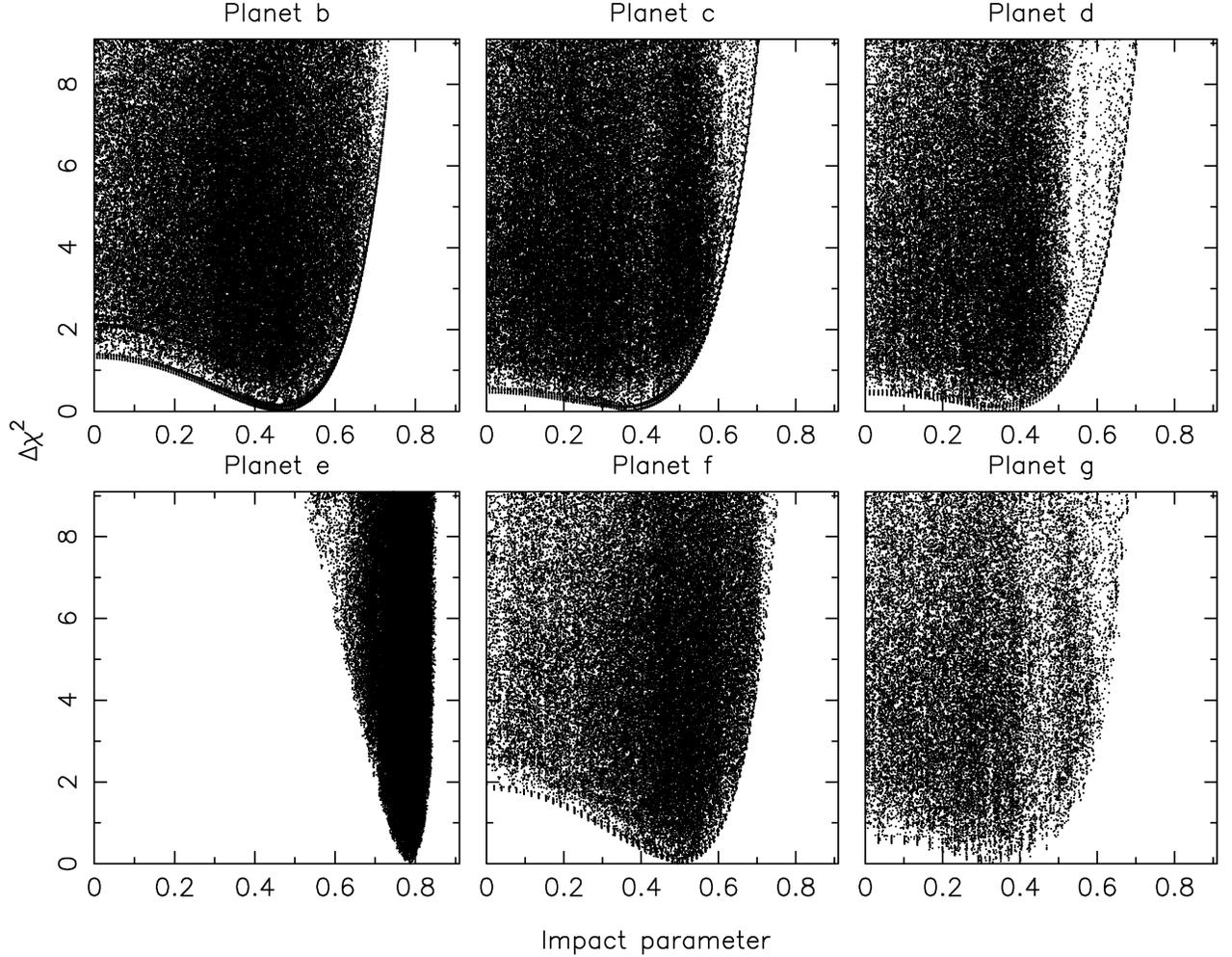

Fig. S4.— $\Delta\chi^2$ as a function of the impact parameter $b$ from the model fits performed using a Monte Carlo Markov Chain. The nominal $1\sigma$ confidence limits are taken to be the parameter range where $\Delta\chi^2 < 1$. Planet Kepler-11e has a non-zero impact parameter at a high significance. Planets Kepler-11f and b also have non-zero impact parameters, although in these cases the significance is less than $2\sigma$. Planets Kepler-11c, d, and g have impact parameters consistent with zero at the $1\sigma$ level.

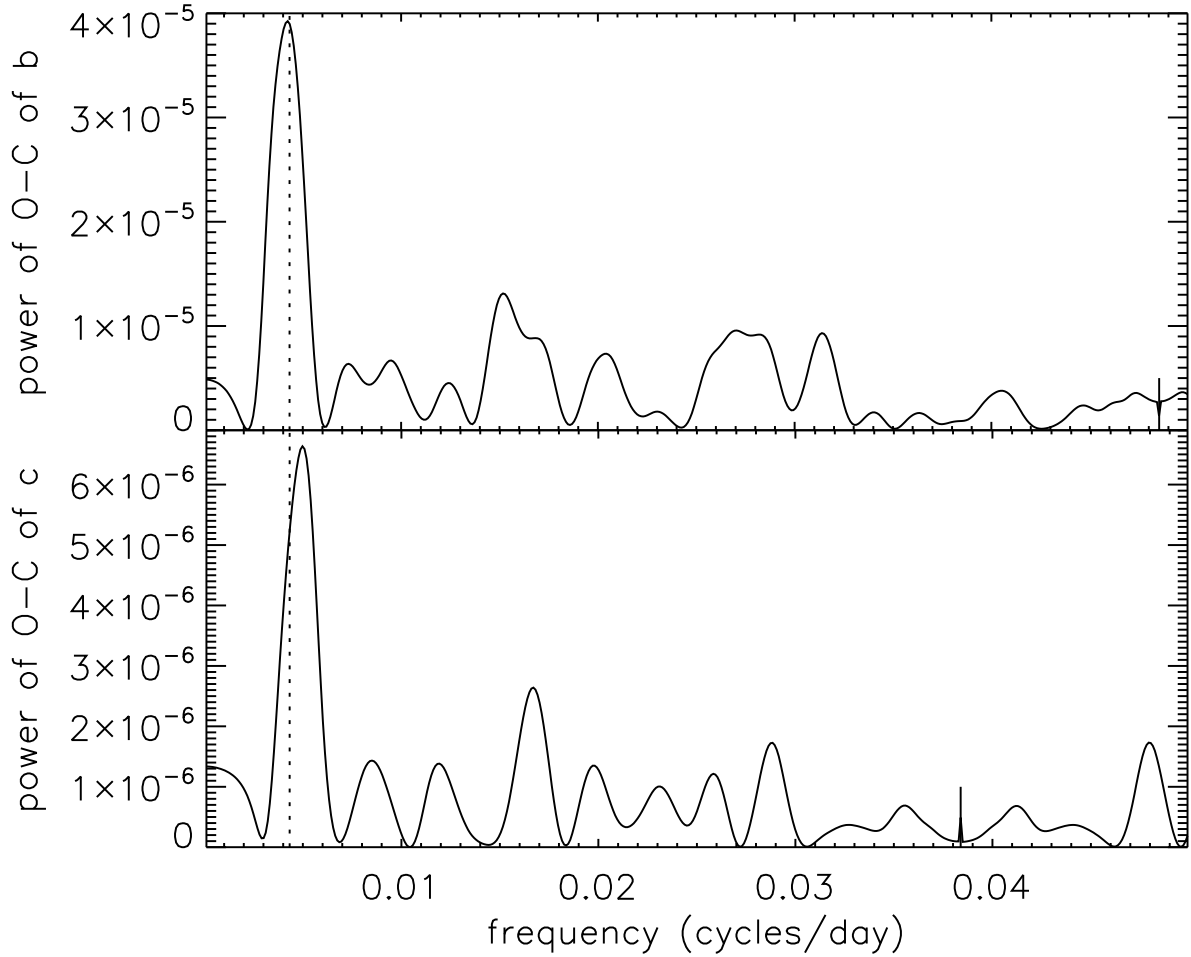

Fig. S5.— The spectral power of the $O - C$ data (Figure 3, main text) for planets Kepler-11b and c. The expected peak frequency caused by the deviation of the planets from the nearby 4:5 mean motion resonance is shown as a dashed line. Large tick marks on the right sides are the Nyquist frequency for each planet, beyond which the spectrum holds no additional information.

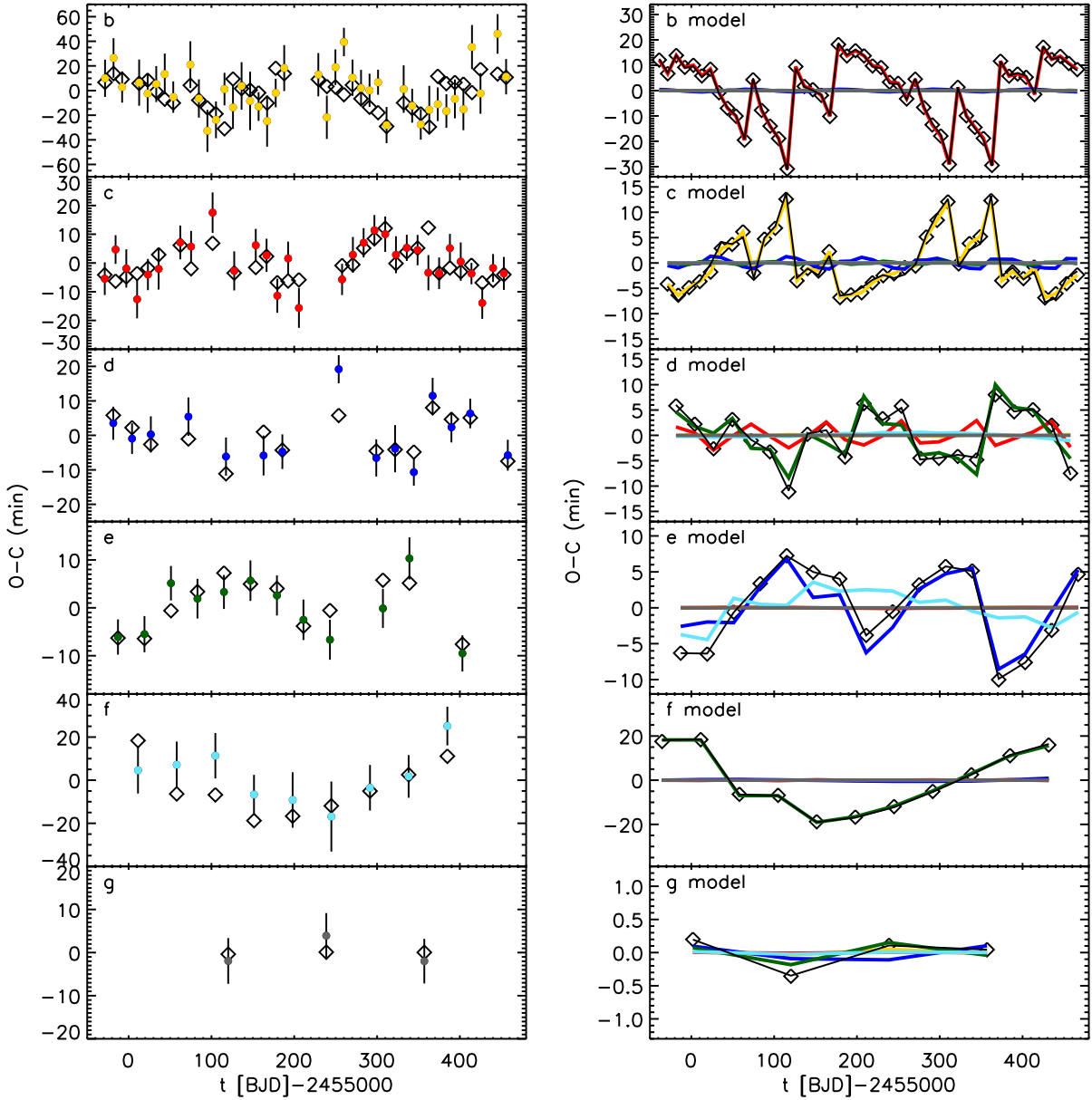

Fig. S6a.— Left side: Observed (O) mid-times of planetary transits minus a Calculated (C) linear ephemeris, plotted as dots with error bars; colors correspond to Figures 1–3 in the main text. Numerical integration dynamical model, the Circular Fit of Table S4, is given by the open diamonds. Right side: Contributions of individual planets to these variations. Total variations from saw ix-planet integrations are given as diamonds (same values but different scale than left side), and contributions from every other planet is shown by a line with color corresponding to the perturbing planet, determined by two-planet integrations. The solid black line is the sum of these integrations, which matches nearly identically with the diamonds; thus we conclude the perturbations from different planets add up very linearly.

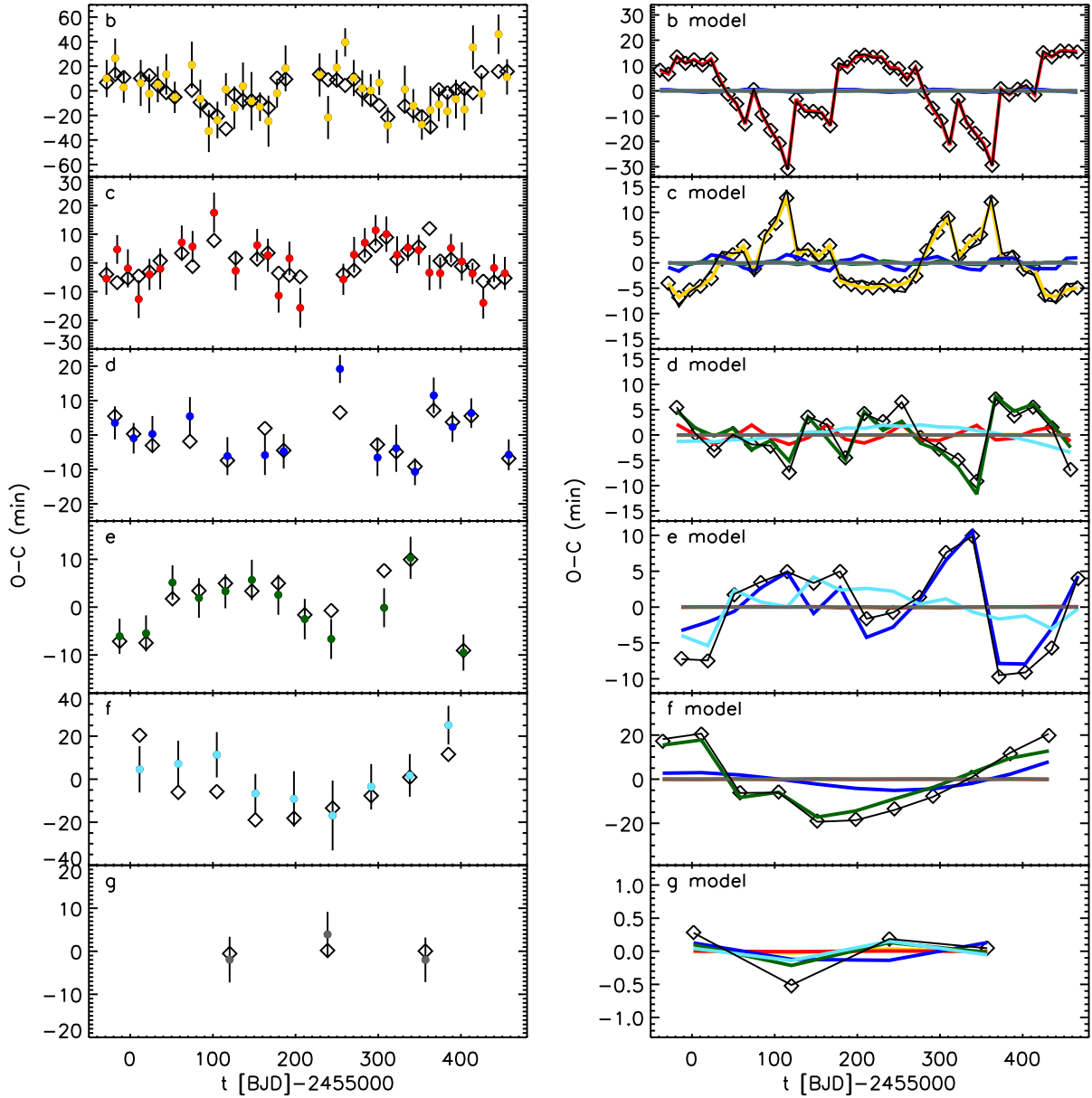

Fig. S6b.— Same as Figure S6a, but using the all-eccentric fit of Table S4. Most features are similar. The b/c-eccentric fit of Table S4 produces a figure which is almost visually identical to this one.

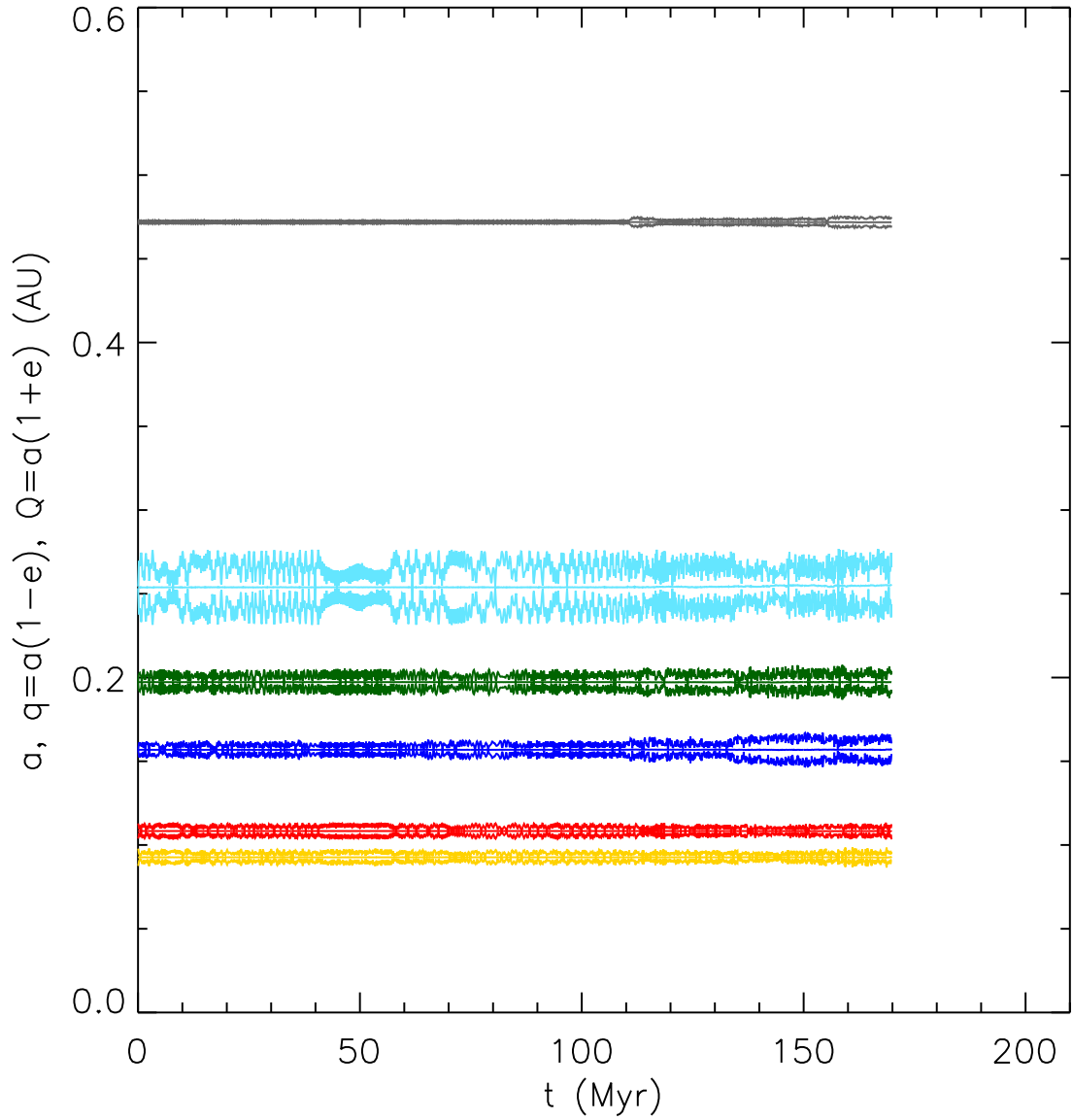

Fig. S7.— Instability of a system that fits the transit times (b/c-eccentric fit of Table S4). The eccentricity variations are chaotic, and a system with almost identical initial conditions survived at least 30% longer.

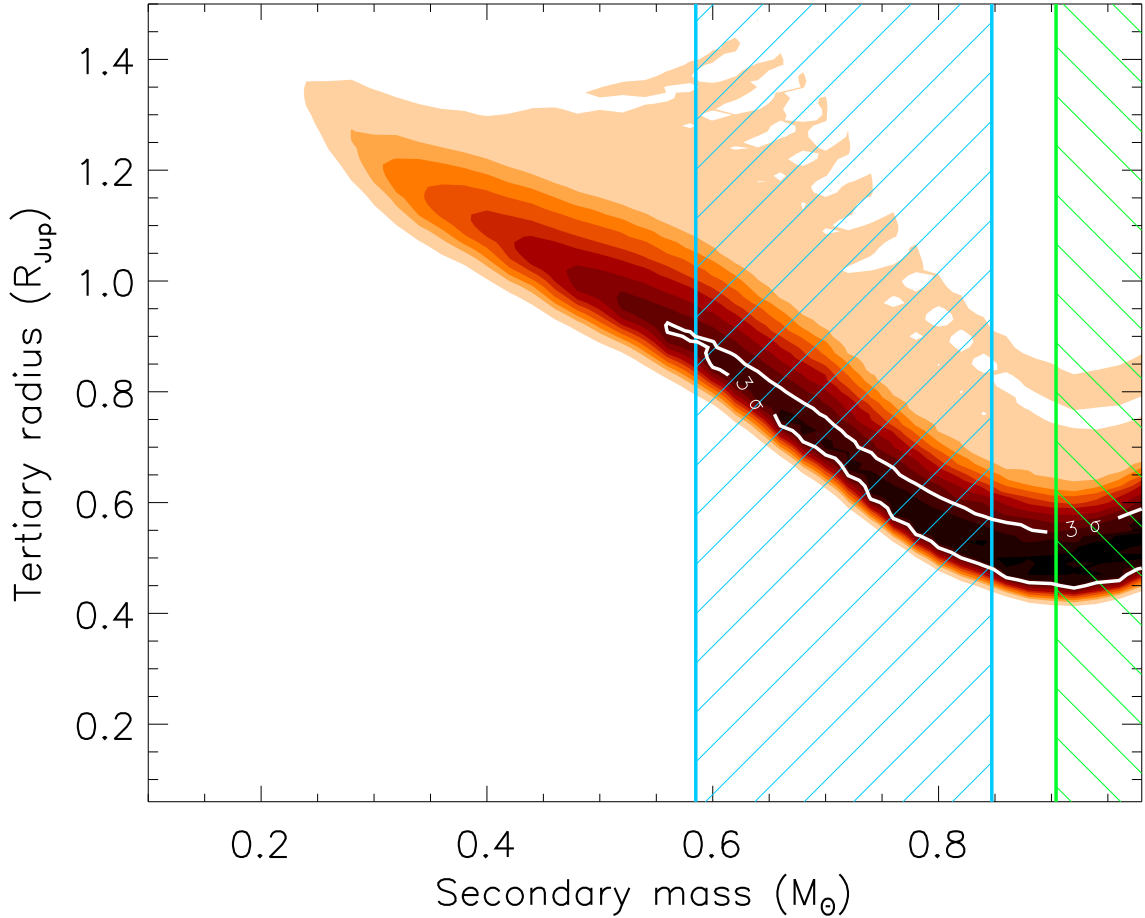

Fig. S8.— Map of the $\chi^2$ surface (goodness of fit) corresponding to a grid of blend models for Kepler-11g involving hierarchical triple systems in which the tertiary is a giant planet. The contours shown represent fixed values of the $\chi^2$ difference from the best planet model fit, expressed in units of the significance level of the difference, $\sigma$. Only the 3-$\sigma$ (white) contour is labeled, for clarity. The green vertical line is drawn at the largest mass for the secondary ($0.91\,\mathrm{M_\odot}$) that would be faint enough to be missed spectroscopically ($\Delta Kp = 1$). Blends to the right (hatched area) are brighter and are ruled out. The smallest secondary mass that still provides an acceptable fit to the *Kepler* light curve ($0.55\,\mathrm{M_\odot}$) corresponds to a brightness difference in the *Kepler* band of about 3.5 magnitudes relative to the target. The hatched region between the vertical blue lines represents the area of parameter space for blends that are too red compared to the measured $r - K_s$ color of Kepler-11 ($3\sigma$ difference of 0.11 mag), and is therefore excluded. Consequently, the only blends that cannot be ruled out by any follow-up observations are those within the white $3\sigma$ contour that have secondary masses between 0.55 and $0.58\,\mathrm{M_\odot}$, or between 0.85 and $0.91\,\mathrm{M_\odot}$. The tertiaries in these blends are roughly $0.5\,\mathrm{R}_J$ or $\sim 1\,\mathrm{R}_J$ in size, respectively.

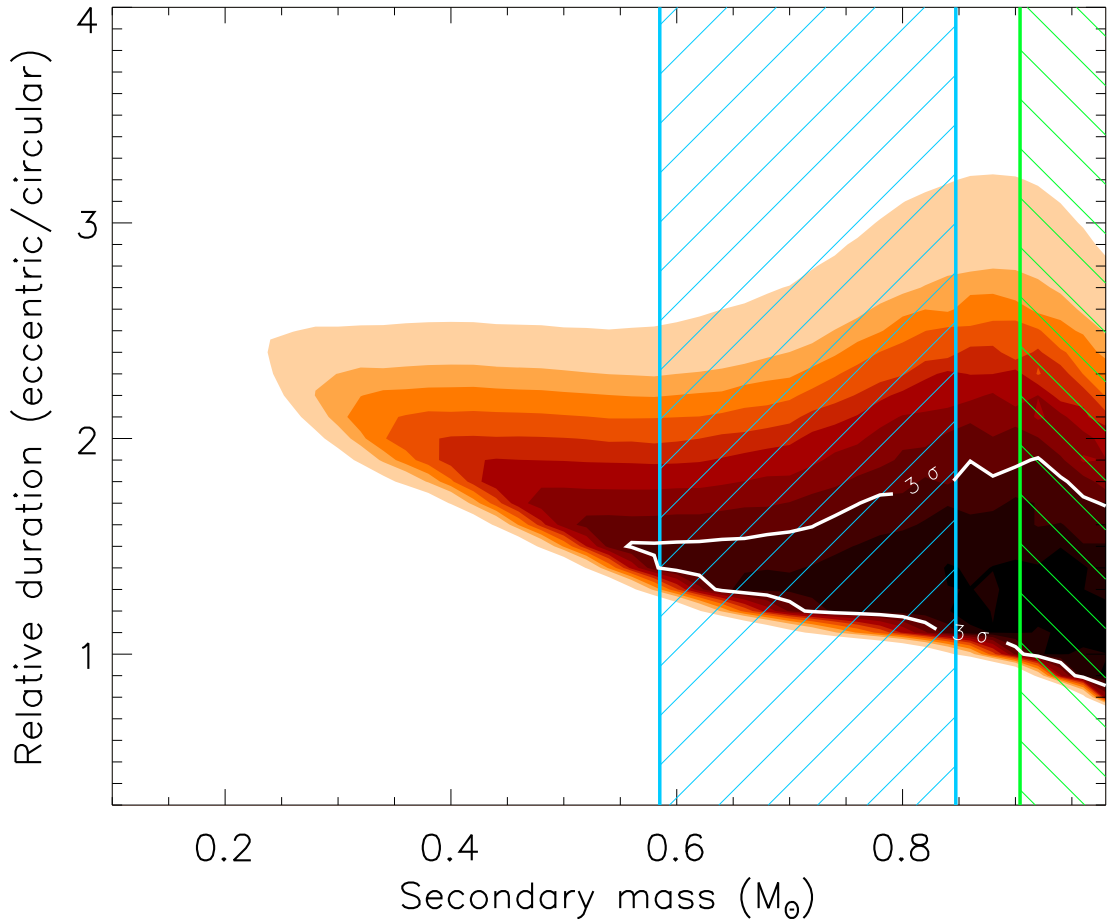

Fig. S9.— Similar to Figure S8 (blends involving hierarchical triple systems in which the tertiary is a giant planet), showing the duration of the predicted transits relative to the duration for a circular orbit as a function of secondary mass.

– 35 –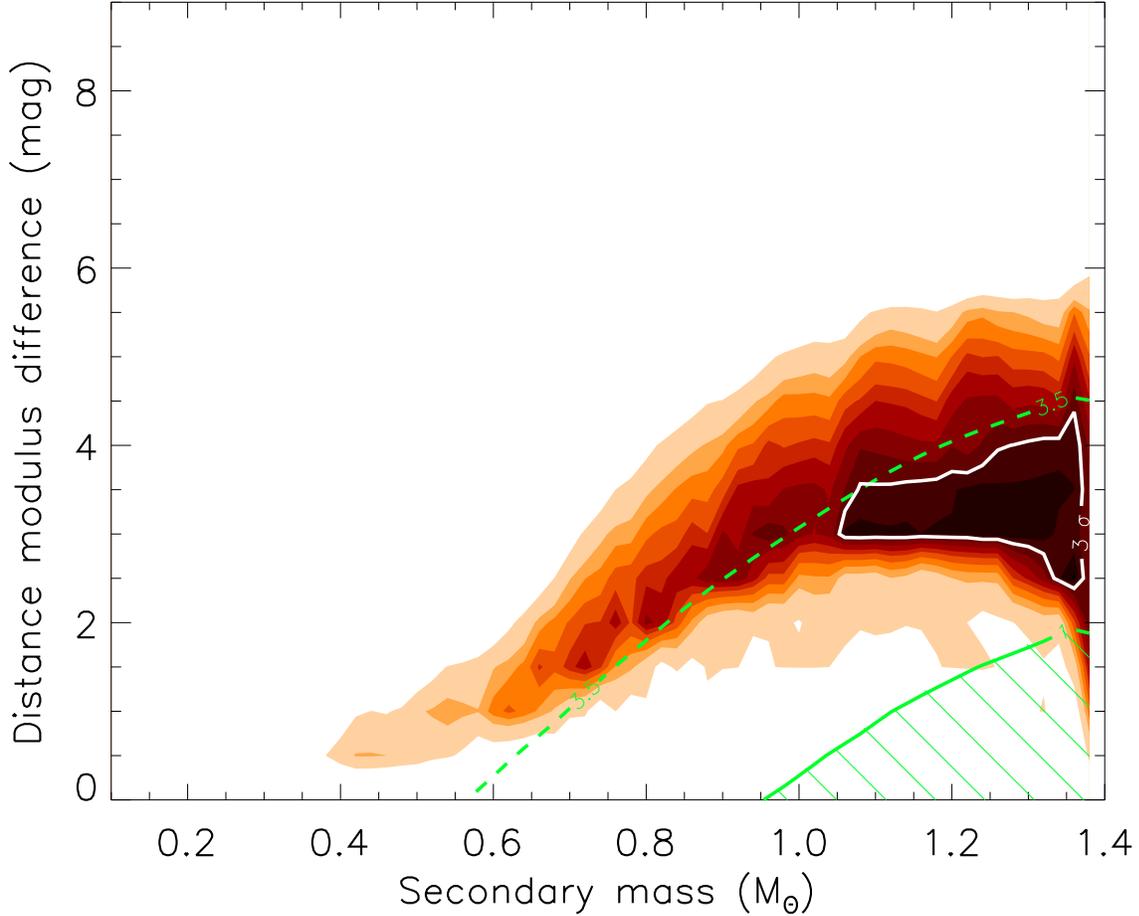

Fig. S10.— Map of the $\chi^2$ surface corresponding to a grid of blend models for Kepler-11g for the case of a background eclipsing binary (star+star). The vertical axis shows the relative distance between the binary and the main star expressed in terms of the difference in the distance modulus. The solid green line corresponds to a brightness difference of $\Delta Kp = 1$ between the target and the background binary. Blends brighter than this (hatched area) are ruled out as they would have been detected spectroscopically. The faintest viable blends have $\Delta Kp = 3.5$ (dashed green line).

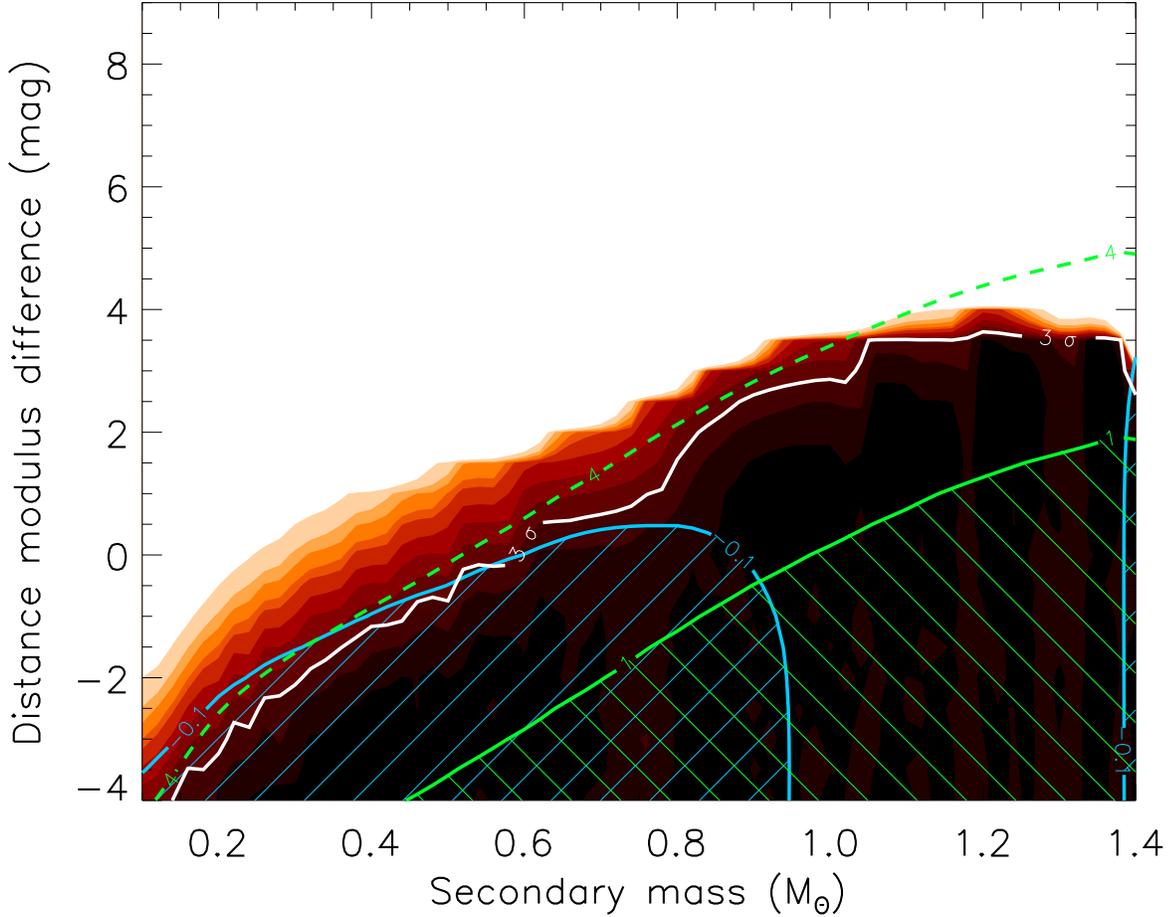

Fig. S11.— Similar to Figure S10, but for the case of blends consisting of an eclipsing star+planet pair. Note that many of the configurations that provide good fits to the data are in the foreground (negative distance modulus differences). The hatched area below the blue line corresponds to blends that are significantly redder in $r - K_s$ (by $3\sigma$, or 0.11 mag) compared to the measured color of Kepler-11, and is excluded. The solid green line corresponds to a brightness difference of $\Delta Kp = 1$ between the target and the background/foreground star. Blends brighter than this (hatched area) are ruled out as they would have been detected spectroscopically. Only blends below the white $3\sigma$ contour in regions that are not hatched are permitted by the constraints placed by the observations.

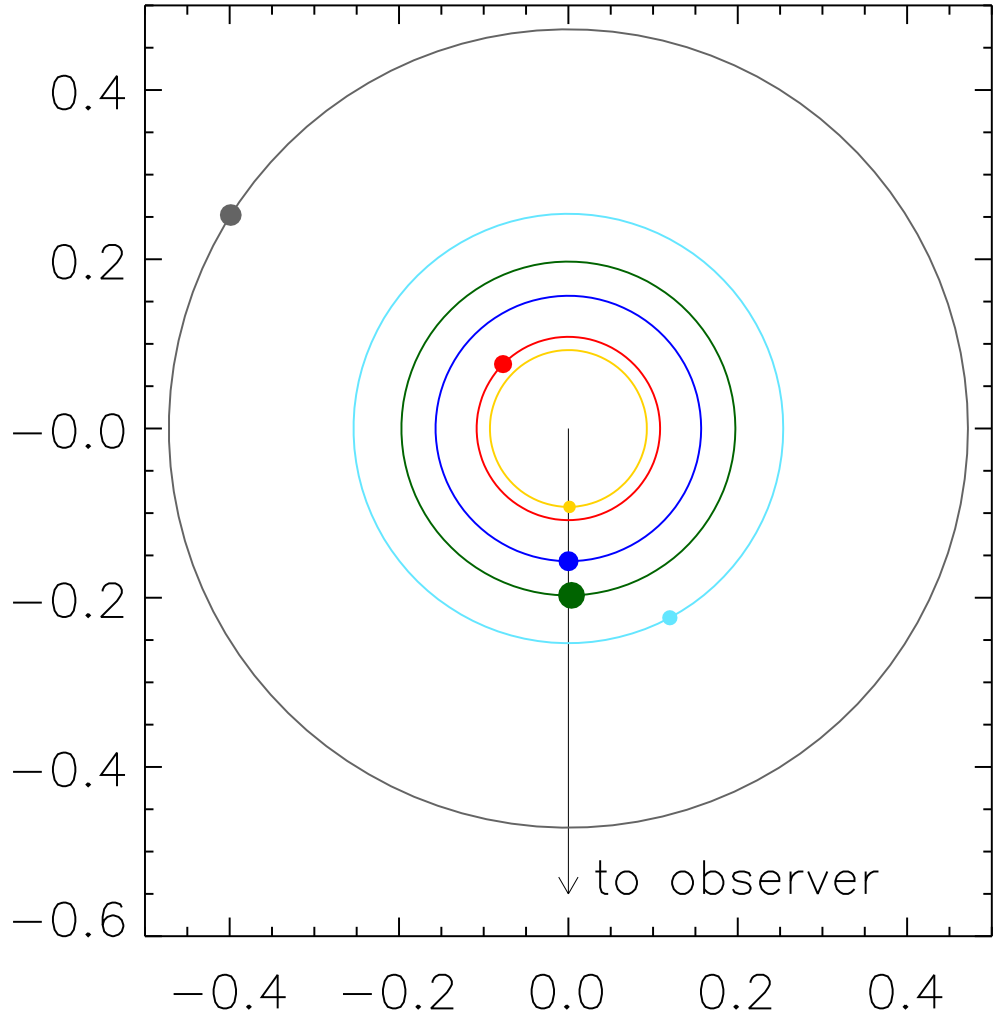

Fig. S12.— Planetary configuration during the triple transit seen at BJD 2,455,435.2. The radii of the points are scaled to the radius of each planet. Orbits are also to scale with one another, but planetary radii are exaggerated relative to orbital ones for clarity. Planetary colors match Figures 1 and 2 in the main text.